\documentclass[a4paper,11pt]{article}
\pdfoutput=1 

\usepackage{jheppub}

\usepackage[utf8]{inputenc}
\usepackage{float} 
\usepackage{graphicx}
\usepackage{caption}
\usepackage{subcaption}
\usepackage{dcolumn}
\usepackage{bm}
\usepackage[nameinlink]{cleveref}
\usepackage[usenames,dvipsnames]{xcolor}
\usepackage[normalem]{ulem}
\usepackage{amsmath}
\usepackage{booktabs}
\usepackage{siunitx}
\definecolor{rossoCP3}{cmyk}{0,.88,.77,.40}
\definecolor{blaa}{rgb}{0.2,0.2,0.6}

\newcommand{\radd}[1]{#1}

\title{Conformal versus non-conformal two-Higgs-doublet model: phase transitions and gravitational waves}

\author[a,1]{Nico Benincasa}
\author[a,1]{Ji-Wei Li}
\author[b,d,1]{Hanxiao Pu \note{In keeping with common practice in high-energy physics, authors are listed alphabetically by surname. Nico Benincasa, Ji-Wei Li and Hanxiao Pu contributed equally and share first authorship.}}
\author[b]{Robert B. Mann}
\author[c]{Vahid Shokrollahic}
\author[c]{T.G. Steele}
\author[a,2]{Zhi-Wei Wang \note{Corresponding author.}}
\emailAdd{zhiwei.wang@uestc.edu.cn,\,Tom.Steele@usask.ca,\,rbmann@uwaterloo.ca}
\emailAdd{h3pu@uwaterloo.ca,\,jiweili@std.uestc.edu.cn,\,nico.benincasa@kbfi.ee}

\affiliation[a]{School of Physics, University of Electronic Science and Technology of China, No.~2006, Xiyuan Avenue, West Hi-Tech Zone, Chengdu, China}
\affiliation[b]{Department of Physics, University of Waterloo, Waterloo, ON N2L 3G1, Canada}
\affiliation[c]{Department of Physics and Engineering Physics, University of Saskatchewan, Saskatoon, SK, S7N 5E2, Canada}
\affiliation[d]{Department of Astronomy and Theoretical Physics, Lund University, 22100 Lund, Sweden}

\abstract{In this work we investigate the CP-conserving two-Higgs-doublet model (2HDM) in two realizations: a classically conformal setup (C2HDM) and a non-conformal setup with explicit tree-level quadratic mass terms (NC2HDM). Imposing current theoretical and experimental constraints, we scan the parameter space and analyse the electroweak first-order phase-transition dynamics from the finite-temperature effective potential, determining the relevant thermodynamic scales and the associated parameters $\alpha$ and $\beta/H_*$. 
In the resulting $(\alpha,\beta/H_*)$ phase diagrams, the NC2HDM spans a substantially broader region and hosts the strongest transitions, whereas the C2HDM is confined to a nested, weaker-transition subset.
This challenges the common expectation that classical conformal symmetry generically implies deep supercooling. By relaxing the Higgs-mass identification and varying the scalon mass, we show that sizable supercooling is obtained only when the radiative (one-loop) breaking of scale invariance is sufficiently mild, i.e.\ for a light scalon. We then compute the resulting stochastic gravitational-wave spectra and show that only the NC2HDM yields benchmark points potentially observable by future space-based interferometers such as LISA, TianQin and Taiji (and, in favourable cases, by more sensitive missions such as DECIGO/BBO).}

\begin{document}
\maketitle

\section{Introduction}





The Higgs boson, the only elementary scalar particle in the Standard Model (SM), was discovered at the LHC in 2012 by the ATLAS~\cite{ATLAS:2012yve} and CMS~\cite{CMS:2012qbp} experiments, completing the particle content predicted by the SM. In the SM, the Higgs field is an ${\rm SU}(2)_L$ doublet charged under the electroweak (EW) gauge group ${\rm SU}(2)_L\times{\rm U}(1)_Y$. Nevertheless, the SM must be extended to address open questions such as the origin of the baryon asymmetry of the universe and the nature of dark matter. A simple and well-motivated extension is to add a second ${\rm SU}(2)_L$ scalar doublet~\cite{Lee:1973iz,Deshpande:1977rw,Bittar:2025lcr}. 
This beyond-the-SM scenario is known as the two-Higgs-doublet model (2HDM).

Historically, the idea of extending the SM Higgs sector to two scalar doublets predates supersymmetry: it was introduced by T.\,D.~Lee in the context of spontaneous $T$ (equivalently CP) violation, where a second Higgs doublet provides the minimal scalar content required for such a mechanism~\cite{Lee:1973iz,Lee:1974fj}. Later, two-Higgs-doublet structures also arise naturally in supersymmetric constructions. In particular,
the minimal supersymmetric extension of the SM employs two Higgs doublet superfields (today's $H_u$ and $H_d$) to generate fermion masses consistently with supersymmetry, as already realized in the early SUSY
electroweak models~\cite{Fayet:1976et,Fayet:1977yc} (see also the standard review~\cite{Haber:1984rc}). 
In the CP-conserving 2HDM that we shall consider, the two doublets give rise to five physical scalar states: two CP-even scalars, one CP-odd scalar, and a charged pair. A comprehensive review of 2HDMs can be found in Ref.~\cite{Branco_2012}.
Extending the scalar sector also modifies the finite-temperature dynamics of EW symmetry breaking: while the EW phase transition is a crossover in the SM~\cite{Kajantie:1996mn}, the 2HDM can accommodate a first-order phase transition (FOPT) instead. Such a strong EW FOPT is a necessary (though not sufficient) ingredient for electroweak baryogenesis, and it can also generate a stochastic GW background, as discussed below.

Cosmological first-order phase transitions in the early universe proceed through the nucleation, expansion, and eventual collision of bubbles of the broken phase. The resulting anisotropic stress in the energy--momentum tensor can source a stochastic gravitational-wave (GW) background~\cite{Witten:1984rs,Hogan:1986dsh}. Gravitational waves were detected for the first time in 2015~\cite{LIGOScientific:2016aoc}. A key feature of GWs is that they propagate essentially freely, being redshifted only by the expansion of the universe. They therefore provide a powerful probe of the earliest moments of cosmic history---complementary to collider searches---since a stochastic signal directly encodes information on the finite-temperature effective potential and the thermal history of symmetry breaking~\cite{Caprini:2018mtu,Auclair:2022wcv}.

Moreover, for causal sources such as phase transitions (characterized by a strength parameter $\alpha$), the characteristic GW frequency observed today scales approximately with the transition temperature (and with $\beta/H_*$, the ratio of the inverse time duration of the phase transition to the Hubble parameter), so that different GW bands map to different energy scales: pulsar timing arrays (nHz) can probe low-temperature transitions at the MeV--GeV scale~\cite{Arzoumanian:2021fsy,Xue:2021gyq}, space-based interferometers (mHz) target electroweak- and TeV-scale transitions, while extending searches to higher frequencies opens a window on much higher-scale transitions that are far beyond the reach of direct collider production~\cite{Huang:2020bbe,Aggarwal:2021icw,Guo:2025uhf}. A detection of a GW background from an EW-scale transition would provide strong evidence for new physics in the scalar sector, while null results can place meaningful constraints on such scenarios. In the near future, space-based GW observatories such as LISA~\cite{2017arXiv170200786A}, TianQin~\cite{TianQin:2015yph}, and Taiji~\cite{Hu:2017mde}, operating in the mHz band, will therefore provide a particularly sensitive probe of electroweak-scale phase transitions.

Motivated by the prospect of probing electroweak symmetry breaking with future GW measurements, it is timely to examine theoretically well-motivated extensions of the Higgs sector that can accommodate a strong first-order phase transition. 
Beyond purely phenomenological considerations, it is also instructive to consider extensions of the Higgs sector in which the tree-level scalar potential is scale invariant, i.e.\ it contains only quartic couplings and no explicit mass parameters.
In such setups, electroweak symmetry breaking may be triggered radiatively through \emph{perturbative dynamical symmetry breaking}: loop corrections lift a flat direction and generate a physical scale by \emph{dimensional transmutation}, as in the Coleman--Weinberg mechanism and its multi-scalar generalisation by Gildener and Weinberg~\cite{Coleman:1973jx,Gildener:1976ih}. Classical scale invariance has also been advocated as a possible ingredient to improve the naturalness of the electroweak scale, but the extent to which it addresses the hierarchy problem depends on additional assumptions about ultraviolet completion (in particular the absence of intermediate heavy thresholds) and remains an active subject of discussion~\cite{Bardeen:1995kv,Meissner:2006zh,Meissner:2007ru,Shaposhnikov:2008xi,Foot:2007as,Steele:2012av,Steele:2013fka,Steele:2014dsa,Hill:2014mqa,Chataignier:2018kay}. Interestingly, Coleman--Weinberg symmetry breaking can be compatible with complete asymptotic freedom, as explored in Ref.~\cite{Hansen:2017pwe}. In this framework, the pseudo-Goldstone boson of broken scale symmetry (often referred to as the scalon) acquires a mass at one loop, and its mass provides a convenient measure of the radiative breaking of scale invariance (see Section~\ref{sec:2HDM}, and Section~\ref{sec:limited_supercooling} for its impact on supercooling).

At finite temperature, Coleman--Weinberg-type potentials can lead to particularly strong first-order phase transitions and, in parts of parameter space, to a vacuum-dominated supercooling stage. Such dynamics have been investigated in a variety of classically conformal and nearly conformal extensions of the SM, including conformal $U(1)_{B-L}$ models and related scenarios~\cite{Konstandin:2011dr,Jinno:2016knw,Marzola:2017jzl,Marzo:2018nov,Iso:2017rjs,Ellis:2019oqb}. \radd{The mechanism behind this expectation --- and the reason why it is truncated once the scalon is as heavy as the observed Higgs boson --- is reviewed in Appendix~\ref{app:supercooling}.}  Here we use the 2HDM as a controlled testbed: we compare the classically conformal limit (C2HDM), where the electroweak scale is generated radiatively, with a non-conformal realization (NC2HDM) featuring explicit quadratic mass terms at tree level. This comparison cleanly isolates the role of explicit mass scales in shaping the vacuum structure, the phase-transition strength, and the resulting gravitational-wave signatures. \radd{With respect to the existing 2HDM phase-transition literature --- in particular the CP-conserving analysis of Ref.~\cite{Basler_2017}, whose finite-temperature formalism we adopt for the non-conformal baseline --- the new elements of this work are: the finite-temperature analysis of the classically conformal (Gildener--Weinberg) realization; the side-by-side comparison of the conformal and non-conformal models under a common scan strategy and a common set of theoretical and experimental constraints; the use of percolation-based quantities, which remain well defined even when the nucleation temperature does not exist, as input for the phase-transition parameters and the gravitational-wave spectra; the scalon-mass scan of Section~\ref{sec:limited_supercooling}, which shows that classical scale invariance by itself does not guarantee deep supercooling; and the explicit exponential relation between the transition strength and the scale-breaking coefficient $\mathcal{B}=m_h^2/(8v^2)$ derived in Appendix~\ref{app:supercooling}, $\alpha(\mathcal{B})\simeq[15\mathcal{B}/(\pi^2g_*)]\,e^{4\kappa/\mathcal{B}}$, which quantifies this statement and, to the best of our knowledge, had not been presented before. Part of the analytic ingredients entering this relation were derived in the model-independent analyses of radiative symmetry breaking of Refs.~\cite{Salvio:2023qgb,Salvio:2023ynn,Levi:2022bzt}}

Our paper is structured to isolate the impact of tree-level scale invariance on the electroweak phase transition and its gravitational-wave (GW) signature in the 2HDM. We first construct a consistent finite-temperature effective potential for both the classically conformal and non-conformal realizations, including one-loop and thermal corrections (Section~\ref{sec:2HDM}), and we perform constrained parameter scans subject to theoretical and experimental bounds (Section~\ref{sec:constraints}). We then map the surviving points onto the phase-transition parameter space and demonstrate, via the $(\alpha,\beta/H_*)$ phase diagrams, that the non-conformal model accommodates the strongest transitions, challenging the expectation that classical conformal symmetry generically implies deep supercooling (Section~\ref{sec:phase_transition}). Furthermore, by relaxing the Higgs-mass identification we show that sizable supercooling in the conformal model arises only when the radiative (one-loop) breaking of scale invariance is sufficiently weak, i.e.\ for a light scalon (Section~\ref{sec:limited_supercooling}). Finally, using percolation-based inputs we predict the resulting stochastic GW spectra and assess their detectability against the sensitivities of upcoming space-based interferometers (Section~\ref{sec:gravitational_waves}). We conclude in Section~\ref{sec:conclusion}.

\section{Model setup: conformal vs non-conformal two-Higgs-doublet model}
\label{sec:2HDM}
In this section, we analyse the finite-temperature effective potential in both classically conformal and non-conformal two-Higgs-doublet models. Our analysis is based on the CP-conserving 2HDM framework presented in \cite{Basler_2017}. \radd{Specifically, we take from Ref.~\cite{Basler_2017} the field conventions, the thermal masses and the one-loop counterterm prescription employed for the non-conformal model; the conformal construction discussed below, the common-constraint comparison of the two realizations, and the percolation-based phase-transition and gravitational-wave analysis of Sections~\ref{sec:phase_transition} and~\ref{sec:gravitational_waves} are new to the present work.} We begin by discussing the conformal scenario, followed by the non-conformal case.

\subsection{Conformal 2HDM: setup and effective potential}

In this subsection, we focus on the classically conformal two-Higgs-doublet model. We begin with the tree-level potential, then incorporate the one-loop Coleman–Weinberg corrections, and finally include the finite-temperature contributions.
\subsubsection{Tree-level potential, flat direction and spectrum}

Phenomenological studies of the 2HDM often rely on simplifying assumptions. To suppress potentially large flavor-changing-neutral-currents (FCNCs), a discrete \( \mathbb{Z}_2 \) symmetry is typically imposed, such as $\Phi_1\rightarrow \Phi_1,\,\Phi_2\rightarrow -\Phi_2$,  where the fields \( \Phi_i \) denote complex scalar SU(2)\(_L\) doublets. As summarized in Table~\ref{table:1}, the 2HDM fermion sector is commonly classified into type-I, type-II, type-X and type-Y models, based on the coupling interactions between right-handed fermions and the Higgs doublets~\cite{Branco:2011iw}. However, following the results in~\cite{Arcadi:2022lpp}, we can safely ignore the distinctions between different 2HDM types, as our analysis focuses exclusively on the top quark. The contributions from lighter fermions are negligible in the context of phase transition dynamics due to their small masses.
\begin{table}[H]
\centering
\begin{tabular}{|c|c|c|c|c|}
\hline
~~~~~~ &  Type I & Type II & Type X & Type Y \\ \hline  
$u_R^i$ & $\Phi_2$ & $\Phi_2$ & $\Phi_2$ & $\Phi_2$ \\ \hline
$d_R^i$ & $\Phi_2$ & $\Phi_1$ & $\Phi_2$ & $\Phi_1$ \\ \hline
$e_R^i$ & $\Phi_2$ & $\Phi_1$ & $\Phi_1$ & $\Phi_2$ \\ \hline
\end{tabular}
\caption{Classification of 2HDM types depending on the interactions between $\Phi_i$ and right-handed fermions.}
\label{table:1}
\end{table}

Assuming CP conservation, the tree-level potential of a general classical conformal 2HDM is given by~\cite{Lee_2012,Branco:2011iw}:
\begin{equation}
\begin{aligned}
V_{\text{tree}} = & \frac{\lambda_1}{2}(\Phi_1^\dagger \Phi_1)^2 + \frac{\lambda_2}{2} (\Phi_2^\dagger \Phi_2)^2 + \lambda_3 (\Phi_1^\dagger \Phi_1)(\Phi_2^\dagger \Phi_2) 
\\ &+ \lambda_4 (\Phi_1^\dagger \Phi_2)(\Phi_2^\dagger \Phi_1) + \frac{\lambda_5}{2} \left[(\Phi_1^\dagger \Phi_2)^2 + (\Phi_2^\dagger \Phi_1)^2\right] ,
\end{aligned}
\label{eq:C2HDM_Vtree}
\end{equation}
where all quartic couplings $\lambda_i$ are real and dimensionless.  Assuming that only the neutral components of the Higgs doublets acquire non-vanishing vacuum expectation values (VEVs), they can be parameterized as follows:
\begin{equation}
\label{field}
\Phi_i = \frac{1}{\sqrt{2}}\left( \begin{array}{c}
\sqrt{2} \phi_i^+ \\
 v_i + \rho_i + i \eta_i 
\end{array} \right), \quad i = 1, 2\,,
\end{equation}
where \(\rho_i\) and \(\eta_i\) denote the CP-even and CP-odd neutral components, respectively. After electroweak symmetry breaking---dynamically induced by quantum effects in the C2HDM (see Section~\ref{sec:C2HDM-CW})---the two Higgs doublets acquire VEVs, denoted as \(v_1=\ v\cos\tilde{\beta}\equiv v c_{\tilde{\beta}}\) and \(v_2=\ v\sin\tilde{\beta}\equiv v s_{\tilde{\beta}}\), such that one has \(\sqrt{v_1^2+v_2^2}=v\simeq 246\) GeV. \radd{Throughout this work we restrict the analysis, both at zero and at finite temperature, to electrically neutral background-field configurations, $\langle\phi_i^\pm\rangle=0$: a vacuum with a charged-Higgs expectation value would spontaneously break $U(1)_{\rm em}$ and is not considered. We recall that, at tree level, a charge-breaking stationary point of the 2HDM potential always lies above a normal (neutral) minimum~\cite{Ferreira:2004yd}.}

By extremizing the tree-level potential along the CP-even directions
\begin{equation}
\label{mincon}
\left.\frac{\partial V_{\text {tree}}}{\partial \rho_i}\right|_{\rho_i=v_i} = 0, \quad i=1,2,
\end{equation}
one obtains the following stationary conditions:
\begin{equation}
  \frac{\lambda_1}{\lambda_2}=\tan^4\tilde{\beta}, \quad \quad \sqrt{\lambda_1\lambda_2}=-\lambda_{345}\,,
  \label{flatmin}
\end{equation}
with $\tan\tilde{\beta} = v_2/v_1$  and $\lambda_{345}\equiv\lambda_3+\lambda_4+\lambda_5$.

The mass matrices for the CP-even, CP-odd, and charged Higgs at tree level are obtained by evaluating the second derivatives of the tree-level potential:
\begin{equation}
\label{tmmatrix}
(\mathcal{M})_{i j}=\left.\frac{\partial^2 V_{\text {tree }}}{\partial \varphi_i \partial \varphi_j}\right|_{\varphi=\varphi^c}
\end{equation}
with $\varphi_i \equiv\left\{\rho_1,\rho_2,\eta_1,\eta_2, \phi^+_1,\phi^+_2\right\}$ evaluated at the classical constant field configurations $\varphi_i^c \equiv\left\{\phi_1,\phi_2,0,0,0,0\right\}$. The physical masses are given by the field values in the global minimum of the potential, ($\phi_1,\phi_2) = (v_1,v_2)$.  The mass terms in the tree-level potential can be expressed as~\cite{Lee_2012}
\begin{equation}
\begin{aligned}
V_{\text{mass}}^0 &= \begin{pmatrix} G^+, & H^+ \end{pmatrix} \begin{pmatrix} 0 & 0 \\ 0 & m_{H^\pm}^2 \end{pmatrix} \begin{pmatrix} G^- \\ H^- \end{pmatrix} + \frac{1}{2} \begin{pmatrix} G^0, & A \end{pmatrix} \begin{pmatrix} 0 & 0 \\ 0 & m_A^2 \end{pmatrix} \begin{pmatrix} G^0 \\ A \end{pmatrix} \\
&\quad + \frac{v^2}{2} \begin{pmatrix} \rho_1, & \rho_2 \end{pmatrix} \begin{pmatrix} \lambda_1 c_{\tilde{\beta}}^2 & \lambda_{345} c_{\tilde{\beta}} s_{\tilde{\beta}} \\ \lambda_{345} c_{\tilde{\beta}} s_{\tilde{\beta}} & \lambda_2 s_{\tilde{\beta}}^2 \end{pmatrix} \begin{pmatrix} \rho_1 \\ \rho_2 \end{pmatrix},
\end{aligned}
\end{equation}
where a rotation matrix with angle \(\tilde{\beta}\) has been used to diagonalise the CP-odd and charged mass matrix, and express the corresponding mass term in terms of physical eigenstates. The physical masses are thus given by
\begin{equation}
\begin{aligned}
\begin{pmatrix} G^0 \\ A \end{pmatrix} &= \begin{pmatrix} c_{\tilde{\beta}} & s_{\tilde{\beta}} \\ -s_{\tilde{\beta}} & c_{\tilde{\beta}} \end{pmatrix} \begin{pmatrix} \eta_1 \\ \eta_2 \end{pmatrix}, \quad m_{G^0}^2 = 0, \quad m_A^2 = -\lambda_5 v^2; \\
\begin{pmatrix} G^\pm \\ H^\pm \end{pmatrix} &= \begin{pmatrix} c_{\tilde{\beta}} & s_{\tilde{\beta}} \\ -s_{\tilde{\beta}} & c_{\tilde{\beta}} \end{pmatrix} \begin{pmatrix} \phi_1^\pm \\ \phi_2^\pm \end{pmatrix}, \quad m_{G^\pm}^2 = 0, \quad     m^2_{H^\pm}=-\frac{1}{2}\lambda_{45}v^2,
\end{aligned}
\label{Mass_Charge_CPOdd}
\end{equation}
with $\lambda_{45}\equiv\lambda_4+\lambda_5$.

Regarding the CP-even mass matrix, one diagonalises it by means of a rotation matrix of angle $\tilde{\alpha}$:
\begin{equation}
\begin{gathered}
\begin{pmatrix} \rho_1 \\ \rho_2 \end{pmatrix} = \begin{pmatrix} c_{\tilde{\alpha}} & -s_{\tilde{\alpha}} \\ s_{\tilde{\alpha}} & c_{\tilde{\alpha}} \end{pmatrix} \begin{pmatrix} H \\ h \end{pmatrix}, \\
\begin{pmatrix} c_{\tilde{\alpha}} & s_{\tilde{\alpha}} \\ -s_{\tilde{\alpha}} & c_{\tilde{\alpha}} \end{pmatrix} \begin{pmatrix} \lambda_1 c_{\tilde{\beta}}^2 & \lambda_{345} c_{\tilde{\beta}} s_{\tilde{\beta}} \\ \lambda_{345} c_{\tilde{\beta}} s_{\tilde{\beta}} & \lambda_2 s_{\tilde{\beta}}^2 \end{pmatrix} \begin{pmatrix} c_{\tilde{\alpha}} & -s_{\tilde{\alpha}} \\ s_{\tilde{\alpha}} & c_{\tilde{\alpha}} \end{pmatrix} = \begin{pmatrix} m_H^2/v^2 & 0 \\ 0 & 0 \end{pmatrix},\\
m_H^2 \, = \, -\lambda_{345} \, v^2 \, = \, \sqrt{\lambda_1 \lambda_2} \, v^2, \quad \sin^2(\tilde{\alpha} - \tilde{\beta}) \, = \, 1 \ .
\end{gathered}
\label{eq:zero_mh}
\end{equation}
Using the tadpole conditions Eq.~\eqref{flatmin}, we find that one of the two eigenvalues, i.e.~$m^2_h$, vanishes, indicating a flat direction along $h$ at tree-level where the curvature of the potential is zero. The scalar potential along this direction is therefore radiatively generated at one-loop order (see Section~\ref{sec:C2HDM-CW}). Moreover, as shown in Eq.~\eqref{eq:zero_mh}, the C2HDM naturally resides in the alignment limit, characterised by $\sin^2(\tilde{\alpha} - \tilde{\beta}) \, = \, 1$, or equivalently $\tilde{\alpha}=\tilde{\beta}-\pi/2$. The resulting flat direction $\phi$ along which we will work in the rest of this manuscript can be expressed as
\begin{equation}
    \phi\equiv h =-s_{\tilde{\alpha}}\rho_1+c_{\tilde{\alpha}}\rho_2=c_{\tilde{\beta}}\rho_1+s_{\tilde{\beta}}\rho_2,
    \label{eq:flat_direction}
\end{equation}
with $\phi^2 = \rho_1^2+\rho^2_2$ and $\langle\phi\rangle = \sqrt{v_1^2+v_2^2}=v$.

Applying the flat direction conditions Eq.~\eqref{flatmin} together with the mass eigenvalues given in Eq.~\eqref{Mass_Charge_CPOdd} and Eq.~\eqref{eq:zero_mh}, the C2HDM can be conveniently parametrised in terms of the physical masses $m_H$, $m_A$, $m_{H^\pm}$, the VEV $v$, and the mixing angle $\tilde{\beta}$:
\begin{equation}
\label{mcop1}
\begin{split}
    &v^2\lambda_1=m^2_H\tan^2\tilde{\beta},\quad  v^2\lambda_2=\frac{m^2_H}{\tan^2\tilde{\beta}}, \quad v^2\lambda_3=2m^2_{H^\pm}-m^2_H,\\ &v^2\lambda_4=m^2_A-2m^2_{H^\pm},\quad v^2\lambda_5=-m^2_A\,.
    \end{split}
  \end{equation}

\subsubsection{One-loop effective potential in the C2HDM}
\label{sec:C2HDM-CW}

Quantum corrections at the loop level play a crucial role in the analysis of classically conformal models. In particular, classical scale symmetry is broken by radiative corrections, leading to the generation of a VEV. The Coleman-Weinberg contribution \(V_{\rm CW}\) in the \(\overline{\rm MS}\) scheme is given by \cite{Coleman:1973jx,quiros1999finite}:
\begin{equation}
\label{cwloop}
V_{\mathrm{CW}}(\phi) = \sum_i \frac{n_i}{64 \pi^2} (-1)^{2 s_i} m_i^4(\phi) \left[\log \left(\frac{m_i^2(\phi)}{\mu^2}\right) - c_i\right],
\end{equation}
where the masses depend on the background field \(\phi\) defined in Eq.~\eqref{eq:flat_direction}. The sum in Eq.~\eqref{cwloop} runs over all particle species in the model, with fermion contributions restricted to the top quark, \(s_i\) denotes the spin of the $i$-th particle and \(n_i\) represents its degrees of freedom, given by \cite{Basler_2017}:
\begin{equation}
\label{dofn}
\begin{aligned}
&n_{\Phi^0} = 1, \quad n_{\Phi^{\pm}} = 2,  \quad n_t = 12, \quad n_{W_T} = 4, \quad n_{W_L} = 2, \\
&n_{Z_T} = 2, \quad n_{Z_L} = 1, \quad n_{\gamma_L} = 1,
\end{aligned}
\end{equation}
with $\Phi^0 \equiv h, H, A, G^0$, $\Phi^{ \pm} \equiv H^{\pm}, G^{\pm}$. The transverse and longitudinal polarisation of the gauge bosons are respectively denoted by $T$ and $L$ and the constants \(c_i\) in the $\overline{\rm MS}$ scheme are given by
\begin{equation}
\label{ci}
c_i = \begin{cases}
\frac{1}{2}, & i = W^{\pm}_T, Z_T \\
\frac{3}{2}, & \text{otherwise}
\end{cases}.
\end{equation}
In the C2HDM, the Coleman-Weinberg potential specifically reads
\begin{equation}
\begin{aligned}
    V_{\rm{CW}}&=\frac{1}{64\pi^2}\bigg[m^4_H\bigg(-\frac{3}{2}+\log\frac{m_H^2}{\mu^2}\bigg)+m^4_A\bigg(-\frac{3}{2}+\log\frac{m_A^2}{\mu^2}\bigg)+2m^4_{H\pm}\bigg(-\frac{3}{2}+\log\frac{m_{H\pm}^2}{\mu^2}\bigg)\\
    &+6m^4_W\bigg(-\frac{5}{6}+\log\frac{m_W^2}{\mu^2}\bigg)+3m^4_Z\bigg(-\frac{5}{6}+\log\frac{m_Z^2}{\mu^2}\bigg)-12m^4_t\bigg(-\frac{3}{2}+\log\frac{m_t^2}{\mu^2}\bigg)\bigg]\,,
    \end{aligned}
\end{equation}
where the background field-dependent tree-level masses are given by
\begin{equation}
\begin{split}
&m^2_H=-\lambda_{345}\phi^2,\quad m^2_A=-\lambda_5\phi^2,\quad
m^2_{H^\pm}=-\frac{\lambda_{45}\phi^2}{2},\\ &m^2_W=\frac{g^2\phi^2}{4},\quad
m^2_Z=\frac{(g^2+g'^2)\phi^2}{4},\quad m^2_t=\frac{y^2_t\rho_2^2}{2},
\end{split}
\end{equation}
with $g$ the SU(2)$_L$ gauge coupling, $g^\prime$ the U(1)$_Y$ gauge coupling and $y_t=y_t^\text{SM}/s_{\tilde{\beta}}$ the rescaled top Yukawa coupling with respect to its SM value. 

At tree level, the flat direction contains a massless scalar, 
which we identify here with the conventional Higgs particle. This \textit{scalon} acquires a mass after the breaking of classical conformal symmetry by quantum corrections through the Coleman-Weinberg mechanism. We set the renormalization scale \(\mu\) to the Gildener-Weinberg scale, \(\mu = \Lambda_{\text{GW}}\),  defined by the relation~\cite{Gildener:1976ih}:
\begin{equation}
\ln \frac{\Lambda_{\text{GW}}}{v} \, = \, \frac{\mathcal{A}}{2\mathcal{B}} + \frac{1}{4} \, ,
\end{equation}
with
\begin{equation}
\begin{aligned}
\mathcal{A} = \frac{1}{64\pi^2 v^4} &\left[ m_H^4 \left( -\frac{3}{2} + \log \frac{m_H^2}{v^2} \right) + m_A^4 \left( -\frac{3}{2} + \log \frac{m_A^2}{v^2} \right) + 2m_{H^\pm}^4 \left( -\frac{3}{2} + \log \frac{m_{H^\pm}^2}{v^2} \right) \right. \\
& \left. + 6m_W^4 \left( -\frac{5}{6} + \log \frac{m_W^2}{v^2} \right) + 3m_Z^4 \left( -\frac{5}{6} + \log \frac{m_Z^2}{v^2} \right) - 12m_t^4 \left( -\frac{3}{2} + \log \frac{m_t^2}{v^2} \right) \right], \\
\mathcal{B} = \frac{1}{64\pi^2 v^4} &\left( m_H^4 + m_A^4 + 2m_{H^\pm}^4 + 6m_W^4 + 3m_Z^4 - 12m_t^4 \right),
\end{aligned}
\end{equation}
and where the masses are evaluated in the vacuum: \(m_i\equiv m_i(v)\). Choosing $\Lambda_{\rm{GW}}$ automatically incorporates the minimization condition into the renormalization scale choice, suppressing large logarithms near the vacuum $v$ and producing cleaner one-loop expressions.\footnote{\radd{This is the standard scale choice in Gildener--Weinberg analyses of classically conformal models~\cite{Gildener:1976ih,Lee_2012}: it builds the one-loop minimization condition into the scale choice, so that the electroweak vacuum and the scalon-mass prediction of Eq.~(\ref{Mhcon}) are implemented without further adjustment, and it minimises the logarithms $\log[m_i^2(v)/\mu^2]$ in the field region around the vacuum where the transition dynamics is determined. The residual $\mu$-dependence is then a higher-order effect within our fixed one-loop treatment. An analysis based on the RG-improved effective potential --- which would be essential in strongly supercooled scenarios, where the transition spans a wide range of temperatures and the couplings run significantly during the transition itself, a situation not encountered in our scans --- is left for future work.}}
Once one-loop quantum corrections are taken into account, the CP-even neutral boson receives a nontrivial modification, while the mass eigenvalues of the other Higgs particles remain unchanged from their tree-level values. The scalon mass \(m_h\), as the pseudo-Goldstone boson of broken scale symmetry, is well approximated by the Gildener-Weinberg mass \(M_{\rm GW}\)~\cite{Lee_2012,Gildener:1976ih}:
\begin{equation}
m_{h}^2 \simeq M_{\text{GW}}^2 \equiv 8 \mathcal{B} v^2,
\label{Mhcon}
\end{equation}
where we take \( m_h \simeq 125 \, \text{GeV} \), corresponding to the SM Higgs particle. 

\subsubsection{Finite-temperature effective potential}

The complete finite temperature effective potential incorporates finite temperature corrections, denoted as \(V_{T}\), which are essential for understanding the behavior of the model at non-zero temperatures. These corrections are given by 
\cite{Dolan:1973qd}:
\begin{equation}
\label{vft}
V_T = \sum_{i} n_i \frac{T^4}{2 \pi^2} J_\pm\left(\frac{m^2_i}{T^2}\right),
\end{equation}
where the thermal integrals \(J_\pm\) are defined as
\begin{equation}
J_\pm\left(\frac{m^2_i}{T^2}\right) = \mp \int_{0}^{\infty} dx \, x^2 \log\left[1 \pm e^{-\sqrt{x^2 + m^2_i/T^2}}\right],
\end{equation}
and where \(J_+(J_-)\) applies for fermions (bosons). We approximate these integrals using polynomial expansions as suggested in \cite{Huang_2020}.
Denoting $a \equiv m_i^2 / T^2$, the thermal integrals for bosons and fermions, in the  high-temperature approximation \((a \ll 1)\), are given by:
\begin{equation}
\label{htpoly}
\begin{aligned}
& J_-^H(a)=-\frac{\pi^4}{45}+\frac{\pi^2}{12} a-\frac{\pi}{6} a^{\frac{3}{2}}-\frac{a^2}{32}\left(\log (a)-c_B\right) \\
& J_+^H(a)=-\frac{7 \pi^4}{360}+\frac{\pi^2}{24} a+\frac{a^2}{32}\left(\log (a)-c_F\right),
\end{aligned}
\end{equation}
with $c_B=3 / 2-2 \gamma_E+2 \log (4\pi)$, $c_F=3 / 2-2 \gamma_E+2 \log (\pi)$ and $\gamma_E \approx 0.5772$ the Euler-Mascheroni constant, while the low-temperature expansion \((a\gg 1)\) yields
\begin{equation}
\label{ltpoly}
J_\pm^L(a)=-\sqrt{\frac{\pi}{2}} a^{\frac{3}{4}} e^{-\sqrt{a}}\left(1+\frac{15}{8} a^{-\frac{1}{2}}+\frac{105}{128} a^{-1}\right).
\end{equation}
The complete finite temperature corrections combine high and low-temperature expansions:
\begin{equation}
\label{completeexpan}
\begin{aligned}
& J_-(a)=e^{-\left(\frac{a}{6.3}\right)^4} J_-^H(a)+\left(1-e^{-\left(\frac{a}{6.3}\right)^4}\right) J_\pm^L \\
& J_+(a)=e^{-\left(\frac{a}{3.25}\right)^4} J_+^H(a)+\left(1-e^{-\left(\frac{a}{3.25}\right)^4}\right) J_\pm^L.
\end{aligned}
\end{equation}

Various resummation schemes have been developed for thermal corrections. These include those based on diagrammatic expansions, among which the Parwani scheme~\cite{Parwani:1991gq} is commonly used in the literature. In this scheme, the squared mass is shifted in both $V_\text{CW}$ and $V_T$ as $m_i^2\rightarrow \overline{m}_i^2 = m_i^2 + c_i T^2$, for $i\in \{W_L, Z_L, \gamma_L, \Phi^0, \Phi^\pm \}$ and where only longitudinal gauge bosons and scalars receive thermal masses.
In addition to the SM content, the constants peculiar to the two scalar doublets in the 2HDM (top coupled to $\Phi_2$ only) are given by~\cite{Basler_2017} 
\begin{equation}
\label{var}
\begin{aligned}
& c_{\Phi_1}=\frac{1}{48}\left[12 \lambda_1+8 \lambda_3+4 \lambda_4+3\left(3 g^2+g^{\prime 2}\right)\right], \\
& c_{\Phi_2}=\frac{1}{48}\left[12 \lambda_2+8 \lambda_3+4 \lambda_4+3\left(3 g^2+g^{\prime 2}\right)+12 y_t^2\right] .
\end{aligned}
\end{equation}
\\
The complete finite temperature effective potential of the C2HDM is thus expressed as
\begin{equation}
\label{C2HDMV}
V_\text{eff}^{\text{C2HDM}} = V_{\text{tree}} + V_{\text{CW}} + V_{T}.
\end{equation}
\subsection{Non-conformal two-Higgs-doublet model}
\subsubsection{Tree-level potential and mass matrices}
For the non-conformal two-Higgs-doublet model, we utilize the same field conventions as in the conformal case, as previously defined in Eq.~\eqref{field}. The NC2HDM tree-level potential is then obtained by adding gauge-invariant quadratic mass terms to the C2HDM one defined in Eq.~\eqref{eq:C2HDM_Vtree}~\cite{Basler_2017}:
\begin{equation}
\begin{split}
V_{\text{tree}}^\text{NC} &= m^2_{11}\Phi^\dagger_1\Phi_1 + m^2_{22}\Phi^\dagger_2\Phi_2 - m^2_{12}\left[\Phi^\dagger_1\Phi_2 + \Phi^\dagger_2\Phi_1\right] \\
&\quad + \frac{\lambda_1}{2}(\Phi^\dagger_1\Phi_1)^2 + \frac{\lambda_2}{2}(\Phi^\dagger_2\Phi_2)^2 + \lambda_3(\Phi^\dagger_1\Phi_1)(\Phi^\dagger_2\Phi_2) \\
&\quad + \lambda_4(\Phi^\dagger_1\Phi_2)(\Phi^\dagger_2\Phi_1) + \frac{\lambda_5}{2}\left[(\Phi^\dagger_1\Phi_2)^2 + (\Phi^\dagger_2\Phi_1)^2\right].
\end{split}
\end{equation}
Applying the extremum conditions Eq.~\eqref{mincon}, one obtains the following relations for the mass parameters \(m^2_{11}\) and \(m^2_{22}\):
\begin{equation}
\begin{split}
    &m^2_{11}=m^2_{12}\tan\tilde{\beta}-\frac{v^2\cos^2\tilde{\beta}}{2}\lambda_1-\frac{v^2\sin^2\tilde{\beta}}{2}\lambda_{345},\\
     &m^2_{22}=\frac{m^2_{12}}{\tan\tilde{\beta}}-\frac{v^2\sin^2\tilde{\beta}}{2}\lambda_2-\frac{v^2\cos^2\tilde{\beta}}{2}\lambda_{345}\,.
\end{split}
\end{equation}
In the NC2HDM, the CP-even, CP-odd and charged tree-level scalar mass matrices are respectively given by: 
\begin{equation}
\begin{split}
\tilde{M}^2_{S}&=    \begin{pmatrix}
m^2_{12}\tan\tilde{\beta}+ \lambda_1 v^2\cos^2\tilde{\beta} & -m^2_{12}+ \lambda_{345}v^2\cos\tilde{\beta} \sin\tilde{\beta} \\ 
-m^2_{12}+ \lambda_{345}v^2\cos\tilde{\beta} \sin\tilde{\beta} & \frac{m^2_{12}}{\tan\tilde{\beta}}+ \lambda_2 v^2 \sin^2\tilde{\beta}
\end{pmatrix},\\
\tilde{M}^2_{P}&=    \begin{pmatrix}
m^2_{12}\tan\tilde{\beta}- \lambda_5 v^2\sin^2\tilde{\beta} & -m^2_{12}+ \lambda_5v^2\cos\tilde{\beta} \sin\tilde{\beta} \\ 
-m^2_{12}+ \lambda_5v^2\cos\tilde{\beta} \sin\tilde{\beta} & \frac{m^2_{12}}{\tan\tilde{\beta}}- \lambda_5 v^2 \cos^2\tilde{\beta}
\end{pmatrix},\\
\tilde{M}^2_{C} &=    \begin{pmatrix}
m^2_{12}\tan\tilde{\beta}- \frac{\lambda_4+\lambda_5}{2}v^2\sin^2\tilde{\beta} & -m^2_{12}+ \frac{\lambda_4+\lambda_5}{2}v^2\cos\tilde{\beta} \sin\tilde{\beta} \\ 
-m^2_{12}+ \frac{\lambda_4+\lambda_5}{2}v^2\cos\tilde{\beta} \sin\tilde{\beta} & \frac{m^2_{12}}{\tan\tilde{\beta}}- \frac{\lambda_4+\lambda_5}{2}v^2 \cos^2\tilde{\beta}
\end{pmatrix}.
\end{split}
\label{eq:NC2HDM_mass_matrix}
\end{equation}
The mixing angle \( \tilde{\alpha} \), which can be expressed as
\begin{equation}
    \tan 2\tilde{\alpha} = \frac{2(\tilde{M}^2_{S})_{12}}{(\tilde{M}^2_{S})_{11}-(\tilde{M}^2_{S})_{22}},
\end{equation}
rotates the neutral  CP-even scalars into their mass eigenstates, while \( \tilde{\beta} \) rotates the charged scalars and neutral CP-odd fields into their mass eigenstates. By diagonalising the three mass matrices above, one can rewrite the couplings in terms of the physical parameters. In the alignment limit i.e.~$\tilde{\alpha}=\tilde{\beta}-\pi/2$\footnote{While in the C2HDM, the alignment was naturally obtained along the flat direction, in the case of NC2HDM we however impose the alignment limit for a better comparison with the C2HDM analysis.}, this parametrisation is given by
\begin{equation}
\label{mcoup2}
\begin{split}
   v^2 \lambda_1 &=  m^2_h + m^2_H \tan^2\tilde{\beta} - m^2_{12} \tan\tilde{\beta}(1+\tan^2\tilde{\beta})\\
v^2 \lambda_2 &=  m^2_h + \frac{m^2_H}{\tan^2\tilde{\beta}} - m^2_{12} \frac{1+\tan^2\tilde{\beta}}{\tan^3\tilde{\beta}}\\
v^2\lambda_3 &= m^2_h- m^2_H + 2m^2_{H^{\pm}} - m^2_{12}\frac{1+\tan^2\tilde{\beta}}{\tan\tilde{\beta}}\\
v^2\lambda_4&=m^2_A-2m^2_{H^{\pm}}+m^2_{12}\frac{1+\tan^2\tilde{\beta}}{\tan\tilde{\beta}}\\
v^2\lambda_5&=-m^2_A+m^2_{12}\frac{1+\tan^2\tilde{\beta}}{\tan\tilde{\beta}}.
\end{split}
\end{equation}

The free parameters in the NC2HDM are thus the physical masses, the soft breaking mass term $m_{12}^2$, the mixing angle $\tilde{\beta}$ and the VEV $v$:
\begin{equation}
m_h, m_H, m_A, m_{H^\pm}, m^2_{12}, \tan\tilde{\beta}, v.
\end{equation}

\subsubsection{One-loop renormalization: counterterms and Goldstone IR treatment}

For the NC2HDM, we employ the same Coleman-Weinberg potential as in the conformal case, given by Eq.~\eqref{cwloop}, Eq.~\eqref{dofn} and Eq.~\eqref{ci}, except that for the scalar part, the eigenvalues are obtained through the diagonalisation of the mass matrices in Eq.~\eqref{eq:NC2HDM_mass_matrix}. Consistent with the C2HDM analysis, we consider the radial direction $\phi=\sqrt{\rho_1^2+\rho_2^2}$. 

In the non-conformal case, loop corrections generally alter the values of the vacuum expectation values (VEVs) and the renormalized mass matrix for the CP-even neutral scalar bosons. To maintain the physical VEVs and masses at their tree-level values, we introduce a counterterm potential ~\cite{Zhang_2021,Basler_2017}  
\begin{equation}
\label{cterms}
    V_{\rm CT}=\delta m_{11}^2\frac {\rho_1^2}{2}+ \delta m_{22}^2\frac{ \rho_2^2}{2}+ \frac{\delta \lambda_1}{8} \rho_1^4+\frac{ \delta \lambda_2}{8} \rho_2^4\\+\delta \lambda_{345}\frac{\rho_1^2 \rho_2^2}{4}\,,
\end{equation}
which must satisfy
\begin{equation}
\begin{split}
     \frac{\partial V_{\rm CT}}{\partial \rho_i}&=- \frac{\partial V_{\rm CW}}{\partial \rho_i}\quad \quad i=1,2 ,\\
   \frac{\partial^2 V_{\rm CT}}{\partial \rho_i\partial\rho_j}&=- \frac{\partial^2 V_{\rm CW}}{\partial \rho_i\partial\rho_j}\quad \quad i,j = 1,2,
\end{split}
\end{equation}
in the vacuum $(v_1, v_2)$. Solving this set of equations, the coefficients of the counterterm potential are given by
\begin{equation}
\begin{split}
\label{m11}
\delta m_{11}^2&= -\frac{3}{2v_1}\frac{\partial V_{\rm CW}}{\partial \rho_1}+\frac{1}{2} \frac{\partial^2 V_{CW}}{\partial\rho_1^2}+ \frac{v_2}{2v_1}\frac{\partial^2 V_{\rm CW}}{\partial \rho_1 \partial \rho_2},\\
\delta m_{22}^2&= -\frac{3}{2v_2}\frac{\partial V_{\rm CW}}{\partial \rho_2}+\frac{1}{2} \frac{\partial^2 V_{CW}}{\partial\rho_2^2}+ \frac{v_1}{2v_2}\frac{\partial^2 V_{\rm CW}}{\partial \rho_1 \partial \rho_2},\\
\delta \lambda_1&= \frac{1}{v_1^3}\frac{\partial V_{\rm CW}}{\partial \rho_1}-\frac{1}{v_1^2} \frac{\partial^2 V_{\rm CW}}{\partial\rho_1^2},\\
\delta \lambda_2&= \frac{1}{v_2^3}\frac{\partial V_{\rm CW}}{\partial \rho_2}-\frac{1}{v_2^2} \frac{\partial^2 V_{\rm CW}}{\partial\rho_2^2},\\
\delta \lambda_{345}&=-\frac{1}{v_1v_2}\frac{\partial^2 V_{\rm CW}}{\partial\rho_1 \partial \rho_2},
\end{split}
\end{equation} 
where all derivatives are evaluated at \((\rho_1,\rho_2)=(v_1,v_2)\). 

In Landau gauge the Goldstone masses satisfy \(m_G^2(\rho)=\xi\,m_V^2(\rho)\) with \(\xi=0\), hence \(m_G^2(v_1,v_2)=0\).
The Goldstone contribution to the one-loop vacuum curvature
\(\partial_{\rho_i}\partial_{\rho_j} V_{\rm CW}\) then contains terms
\(\big(\partial_{\rho_i} m_G^2\big)\big(\partial_{\rho_j} m_G^2\big)\,\ln\!\big(m_G^2/\mu^2\big)\)
that develop an IR logarithm at \((\rho_1,\rho_2)=(v_1,v_2)\): the derivatives are finite at the VEVs while \(m_G^2\to 0\).
To obtain IR-safe second derivatives at the vacuum, we follow the standard prescription~\cite{Cline:1996mga,Cline:2011mm} and replace the divergent Goldstone logs by a finite reference set by the physical Higgs mass,
\begin{equation}
\label{eq:IRsafeGoldstone}
\left[\partial_{\rho_i}\partial_{\rho_j} V_{\rm CW}\right]_{G}
= \sum_{G}\frac{n_G}{32\pi^2}
  \frac{\partial m_G^2}{\partial \rho_i}
  \frac{\partial m_G^2}{\partial \rho_j}
  \ln\frac{m_h^2}{\mu^2},
\qquad
G=G_{0,\pm},\;\; i,j\in\{1,2\}\,.
\end{equation}
Here \([\cdots]_G\) denotes the Goldstone contribution, and all derivatives are
evaluated at \((\rho_1,\rho_2)=(v_1,v_2)\). The degeneracies are
\(n_{G_0}=1\) and \(n_{G_\pm}=2\). This replacement regulates the IR
logarithm by the physical Higgs mass \(m_h\); it is applied only at the
vacuum to fix counterterms or zero-temperature Higgs mass parameters and
does not modify \(V_{\rm CW}\) away from the vacuum.

Combining the counterterm potential with the CW potential, one finally obtains the complete loop-level potential:
\begin{equation}
V_{\text{loop}} = V_{\text{CW}}^\text{NC} + V_{\text{CT}}.
\end{equation}
This procedure ensures that the physical positions of the VEVs and the masses remain unchanged by the loop corrections.

\subsubsection{Thermal effective potential in the NC2HDM}
Analogous to the conformal case, the finite temperature potential for the NC2HDM is defined by Eq.~\eqref{vft}. We employ polynomial expansions, as per Eq.~\eqref{completeexpan}, for computing the thermal integrals. The finite temperature masses are evaluated by analysing the eigenvalues of the finite temperature mass matrices, with the 2HDM coefficient $c_i$ given in Eq.~\eqref{var}.

The complete one-loop finite temperature effective potential for the NC2HDM is thus
\begin{equation}
\label{NC2HDMV}
V_\text{eff}^{\text{NC}} = V_{\text{tree}}^\text{NC} + V_{\text{CW}}^\text{NC} + V_{\text{CT}} + V_T^\text{NC}.
\end{equation}

\section{Constraints and scan setup}
\label{sec:constraints}
This section outlines our setup and scan strategy for the C2HDM and the NC2HDM, including the parametrisation in the alignment limit and the renormalized one-loop potential with the Goldstone IR treatment. 
Working perturbatively, we then impose
perturbative unitarity, and bounded-from-below vacuum stability, as well as constraints from direct searches and electroweak precision data, and then analyse the parameter space that survives.

\subsection{Theoretical constraints}
\label{sec:theor_const}

Perturbativity imposes an upper bound on the quartic couplings: $\vert\lambda_i\vert\leq 4\pi$. Then perturbative unitarity imposes an upper bound on the eigenvalues $e_i$ of the $2\rightarrow 2$ scattering matrix: $\vert e_i\vert\leq 8\pi$. In the 2HDM, these eigenvalues are given by~\cite{Becirevic:2015fmu}
\begin{equation}
    \begin{split}
        &e_{1,2} = \lambda_3\pm \lambda_4, \quad e_{3,4} = \lambda_3\pm \lambda_5, \quad e_{5,6} =  \lambda_3+2 \lambda_4\pm 3 \lambda_5,\\
        &e_{7,8} = \frac{\lambda_1+\lambda_2\pm \sqrt{(\lambda_1-\lambda_2)^2+4\lambda_4^2}}{2}, \quad e_{9,10} = \frac{\lambda_1+\lambda_2\pm \sqrt{(\lambda_1-\lambda_2)^2+4\lambda_5^2}}{2},\\
        &e_{11,12} = \frac{3}{2}(\lambda_1+\lambda_2)\pm\sqrt{\frac{9}{4}(\lambda_1-\lambda_2)^2+(2 \lambda_3+\lambda_4)^2}.
    \end{split}
\end{equation}

Finally, the tree-level potential must be bounded from below. This stability condition translates into~\cite{Gunion:2002zf}
\begin{equation}
    \lambda_1>0,\quad\lambda_2>0,\quad \sqrt{\lambda_1\lambda_2}+\lambda_3 +\min\{0, \lambda_4-\vert\lambda_5\vert\}>0,
    \label{eq:bfb}
\end{equation}
where the last condition cannot be satisfied in the C2HDM because of the flat direction condition Eq.~\eqref{flatmin}~\cite{Lee_2012}.

\subsection{Experimental constraints}
\label{sec:exp_const}

Negative LEP2 searches at a center-of-mass energy up to $\sqrt{s} = 209$ GeV imply~\cite{ParticleDataGroup:2020ssz}: 
\begin{equation}
    m_A \gtrsim 90 \text{~GeV}\quad (e^+e^- \rightarrow hA),\quad m_{H^\pm} \gtrsim 80 \text{~GeV}\quad (e^+e^- \rightarrow H^+H^-).
\end{equation}

Contributions to electroweak radiative corrections from new physics are encapsulated in the oblique parameters $S$, $T$ and $U$~\cite{Peskin:1991sw}. The global fit of the SM predictions to the electroweak
precision data yields~\cite{Lu:2022bgw}
\begin{equation}
    \begin{split}
        &\hat{S} = 0.06 \pm 0.10,\quad \hat{T} = 0.11 \pm 0.12,\quad \hat{U} = -0.02 \pm 0.09,\\
       & \text{corr}(\hat{S},\hat{T}) = 0.90,\quad \text{corr}(\hat{S},\hat{U}) = -0.57,\quad \text{corr}(\hat{T},\hat{U}) = -0.82. 
    \end{split}
    \label{STUdata}
\end{equation} 
Contributions from the heavy scalar states to these oblique parameters are~\cite{Baak:2011ze, Ghorbani:2022vtv}
\begin{equation}
    \begin{split}
       &S =  \frac{1}{12\pi}\left[g\left(m_A^2, m_H^2\right) + \log\frac{m_Am_H}{m_{H^\pm}^2}\right],\\
       &T = \frac{1}{16\pi ~m_W^2\sin^2\theta_W} \left[f(m_A^2, m_{H^\pm}^2) + f(m_H^2, m_{H^\pm}^2)-f(m_A^2, m_H^2)\right],\\
       &U = \frac{1}{12\pi}\left[g(m_A^2, m_{H^\pm}^2) + g(m_H^2, m_{H^\pm}^2)-g(m_A^2, m_H^2)\right],
    \end{split}
    \label{eq:STU}
\end{equation}
with
\begin{equation}
    \begin{split}
         & g(x,y) = \begin{cases}
      -\frac{5}{6} + \frac{2xy}{(x-y)^2} + \frac{(x+y)(x^2-4xy+y^2)}{2(x-y)^3}\log\frac{x}{y} & x \neq y\\
      0 & x=y
    \end{cases} ,\\
    & f(x,y)=\begin{cases}
      \frac{x+y}{2} - \frac{xy}{x-y}\log\frac{x}{y} & x \neq y\\
      0 & x=y
    \end{cases}.
    \end{split}
    \label{eq:STU_func}
\end{equation}
From Eq.~\eqref{eq:STU_func}, we see that the new contributions Eq.~\eqref{eq:STU} to $S$, $T$ and $U$ identically vanish in the limit of a degenerate mass spectrum. A large mass splitting between the heavy scalar states is therefore disfavoured. In order to constrain our model, we require that the total contributions $\vert\hat{S}-S\vert$, $\vert\hat{T}-T\vert$ and $\vert\hat{U}-U\vert$ remain within the 95\% joint confidence level.

\subsection{Scan ranges and strategy}
\label{sec:scan}

To analyse   phase transition dynamics in the C2HDM, we scan the parameter space over the following range:
\begin{equation}
m_A\in[90,1000]~\text{GeV},\quad m_{H^\pm}\in[80,1000]~\text{GeV},\quad \tan\tilde{\beta}\in[0.1,50].
\label{Exlimit}
\end{equation}
In the NC2HDM we also scan the independent inputs
\begin{equation}
  m_H \in [m_h,1~\mathrm{TeV}], \qquad |m_{12}| \in [100,1000]~\mathrm{GeV}.
  \label{eq:scan}
\end{equation}
This contrasts with the conformal case, where scale invariance removes the
dimension-two terms and the Gildener--Weinberg conditions relate the CP-even
masses, so that $m_H$ is not a free input. We retain only points that satisfy
the theoretical and experimental constraints in
Sections~\ref{sec:theor_const} and~\ref{sec:exp_const}. \radd{Note that at the neutral electroweak vacuum the curvature of the potential along the physical charged direction is given precisely by the charged-Higgs squared mass $m_{H^\pm}^2$, the orthogonal charged direction being the flat Goldstone one; since the scan takes $m_{H^\pm}\geq 80$~GeV as a positive physical input, the electroweak vacuum of every accepted point is automatically a local minimum with respect to charged fluctuations, i.e.\ it cannot be destabilised towards a charge-breaking configuration $\langle\phi^\pm\rangle\neq0$. Charge-breaking thermal histories are not included in the present analysis.}

\section{Electroweak phase transition analysis}
\label{sec:phase_transition}

In this section we establish and apply our finite–temperature phase–transition toolkit. 
Sec.~\ref{sec:PT_dynamics} defines the thermodynamic scales \(T_c\), \(T_n\), \(T_p\); the bounce actions \(S_3/T\) and \(S_4\); nucleation/percolation criteria; and the strength/time–scale metrics \(\alpha\) and \(\beta/H_*\), noting caveats from supercooling. 
Sec.~\ref{sec:phase_diagrams} charts the phase diagram and contrasts models: the NC2HDM spans a broader, stronger–transition region (larger \(\alpha\), smaller \(\beta/H_*\)), while the C2HDM forms a nested subset due to milder supercooling. 
Sec.~\ref{sec:limited_supercooling} examines supercooling and vacuum domination in the C2HDM, clarifies when percolation fails, and motivates using \(T_p\) for robust gravitational–wave forecasts.

\subsection{Phase transition formalism}
\label{sec:PT_dynamics}

\subsubsection{Bubble Nucleation}

Very early in the universe the effective potential \(V_{\rm eff}(\phi,T)\) has only a single minimum, singling out a unique vacuum state.
As the universe cools, a second local minimum of the effective potential \(V_{\rm eff}(\phi,T)\) appears and becomes the true vacuum below the critical temperature \(T_c\), where \(V_{\rm eff}(\phi_{\rm true},T_c)=V_{\rm eff}(\phi_{\rm false},T_c)\). This initiates tunneling from the false vacuum to the true vacuum. The tunneling rate per unit volume \( \Gamma(T) \) is \cite{Kobakhidze_2017}
\begin{equation}
\label{gamma}
\Gamma(T)=A(T) e^{-S(T)},
\end{equation}
where \(A(T)\) is a dimension–four prefactor encoding fluctuation determinants. 
A convenient starting point is the Euclidean finite–temperature action (general form)
\begin{equation}
\label{eq:S_general}
S_E[\phi;T]=\int_{0}^{1/T}\! d\tau \int d^3x\,
\bigg[\frac{1}{2}(\partial_\tau\phi)^2+\frac{1}{2}(\nabla\phi)^2
+V_{\rm eff}(\phi,T)-V_{\rm eff}(\phi_{\rm false},T)\bigg].
\end{equation}

Two fluctuation regimes are relevant. At sufficiently high temperature, \emph{thermal} fluctuations dominate and the saddle-point solution is \(\tau\)-independent with \(O(3)\) symmetry; the rate takes the form
\(\Gamma\simeq A_{\rm th}(T)\,e^{-S_3(T)/T}\).
At low temperature, \emph{quantum} fluctuations dominate and the relevant saddle-point solution is \(O(4)\)–symmetric, yielding
\(\Gamma\simeq A_{\rm q}\,e^{-S_4(T)}\)
\cite{Linde:1980tt,Linde:1981zj}.
We adopt the minimal–exponent prescription
\begin{equation}
\label{eq:min_prescription}
S(T)\simeq \min\!\left\{\,S_4(T)\,,\,\frac{S_3(T)}{T}\,\right\},
\end{equation}
and define the crossover temperature \(T_x\) implicitly by
\(S_3(T_x)/T_x=S_4(T_x)\).
Strong supercooling can shift the dominant mechanism from thermal to quantum thereby delaying percolation, in which case the percolation temperature \(T_p\) provides the robust scale for observables.
The \( O(3) \)-symmetric solutions are denoted by \( S_3(T) \), which is
\begin{equation}
\label{S3}
S_3(T)=4 \pi \int_0^{\infty} d r ~r^2\left[\frac{1}{2}\left(\frac{d \phi}{d r}\right)^2 +V_{\rm{eff}}(\phi, T) -V_{\rm{eff}}(0, T) \right],
\end{equation}
with equation of motion and boundary conditions given by
\begin{equation}
\frac{d^2 \phi}{d r^2}+\frac{2}{r} \frac{d \phi}{d r}=\frac{\partial V_{\rm{eff}}}{\partial \phi}(\phi, T),
\end{equation}
\begin{equation}
\frac{d \phi}{d r}(0, T)=0, \quad \lim _{r \rightarrow \infty} \phi(r, T)=0,
\end{equation}
where the bounce $\phi(r)$ describes the bubble configuration and is called the bubble profile (instanton solution). The \( O(4) \)-symmetric solutions are denoted by \( S_4(T) \), which reads
\begin{equation}
\label{S4}
S_4(T)=2 \pi^2 \int_0^{\infty} d \tilde{r}~ \tilde{r}^3\left[\frac{1}{2}\left(\frac{d \phi}{d \tilde{r}}\right)^2+V_{\mathrm{eff}}(\phi, T)-V_{\rm{eff}}(0, T)\right],
\end{equation}
with $\tilde{r}=\sqrt{\tau^2+r^2}$. In that case, the equation of motion and boundary conditions read
\begin{equation}
\frac{d^2 \phi}{d \tilde{r}^2}+\frac{3}{\tilde{r}} \frac{d \phi}{d \tilde{r}}=\frac{\partial V_{\mathrm{eff}}}{\partial \phi}, \quad \phi^{\prime}(0)=0, \quad \phi(\infty)=\phi_{\mathrm{false}}.
\end{equation}
We use the {\tt{CosmoTransitions}} package \cite{Wainwright_2012} to compute the bounce via the shooting method.

\subsubsection{Nucleation temperature and percolation temperature}

The nucleation temperature \( T_n \), when the amount of supercooling is small, can be used to describe the temperature at which gravitational waves from FOPT are generated. It is defined as the temperature at which one bubble is nucleated per horizon on average~\cite{Ellis:2018mja}:
\begin{equation}
\label{tn1}
N\left(T_n\right)=\int_{t_c}^{t_n} d t \frac{\Gamma(t)}{H(t)^3}=\int_{T_n}^{T_c} \frac{d T}{T} \frac{\Gamma(T)}{H(T)^4}=1,
\end{equation}
where $\Gamma(T)$ is given in Eq.~\eqref{gamma} and the second equality comes from isentropic expansion of the universe during the radiation era: $dT=-H T dt$.  
As for the Hubble rate $H(T)$, it is given via the Friedmann equation
\begin{equation}
\label{hrate}
H^2=\frac{1}{3 M_{\mathrm{pl}}^2}\left(\rho_R+\rho_V\right),
\end{equation}
with $M_{\mathrm{pl}}=2.435 \times 10^{18} \mathrm{GeV}$ the reduced Planck mass, the radiation energy density $\rho_{R}$ given by
\begin{equation}
\label{dense}
   \rho_R = \sum_i\rho_{R,i}\quad\quad \rho_{R,i}= \begin{cases} \frac{\pi^{2}}{30}n_iT^{4} & \text{for bosons}\\ \frac{7}{8} \frac{\pi^{2}}{30}n_iT^{4} & \text{for fermions}
    \end{cases},
\end{equation}
and with $\rho_{V}$ the vacuum energy density.

Alternatively, $T_{n}$ can also be defined as the temperature at which the bubble-nucleation rate per unit Hubble volume is unity~\cite{Grojean:2006bp}:
\begin{equation}
\label{tn2}
\frac{\Gamma\left(T_n\right)}{H\left(T_n\right)^4}=1.
\end{equation}
Note that the latter definition of \( T_{n} \) usually corresponds to \( S_{3}(T_n)/T_n \approx 140 \) for electroweak phase transitions, a convention that is typically used to calculate \( T_n \). 

Supercooling can significantly affect the dynamics of the phase transition and the conventional use of \( T_{n} \) may not be adequate to describe scenarios with substantial supercooling. \radd{(The origin of prolonged supercooling in classically scale-invariant potentials is reviewed in Appendix~\ref{app:supercooling}.)}
In this case the percolation temperature \( T_p \) is critical in characterising first-order phase transitions. 
This quantity is numerically determined by the condition \( I(T_p) = 0.34 \)~\cite{1971AdPhy..20..325S} where 
\begin{equation}
\label{pro}
P(T)=e^{-I(T)}, \quad I(T)=\frac{4 \pi}{3} \int_T^{T_c} d T^{\prime}\frac{ \Gamma\left(T^{\prime}\right)}{H\left(T^{\prime}\right) {T^\prime}^4}\left(\int_T^{T^{\prime}} d \tilde{T}\frac{v_b}{H(\tilde{T})}\right)^3,
\end{equation}
with $v_b$  the bubble wall velocity. 
The quantity  $P(T)$ is 
 the probability of being in the false vacuum, and $I(T)$  is  the fraction of the universe  that has transitioned ~\cite{PhysRevD.46.2384,M_gevand_2017,Kobakhidze_2017,Cai_2017,Ellis_2019}. Under the assumption that the bubble wall achieves its final velocity very fast, we can set $v_b=1$. 
 
 For substantial supercooling, it is necessary to check that the physical volume of the false vacuum actually decreases at $T_p$. Indeed, for a large enough $\rho_V$, the universe can enter a period of thermal inflation. Consequently we need to make sure that percolation can actually occur. The condition for the false vacuum volume to shrink is given by~\cite{Turner:1992tz, Ellis:2018mja}
 \begin{equation}
 \label{eq:volume_decreases}
     H(T)\left(3+T\frac{dI(T)}{dT}\right)<0.
 \end{equation}
As will be shown in Section~\eqref{sec:limited_supercooling}, supercooling in the 2HDM is necessarily limited, and therefore Eq.~\eqref{eq:volume_decreases} is automatically satisfied.

\subsubsection{Phase-transition parameters: \( \beta/H_* \) and \( \alpha \)}
\label{intro_beta}

\noindent The parameters\footnote{The phase transition parameters $\alpha$ and $\beta$ should not be confused with the mixing angles $\tilde{\alpha}$ and $\tilde{\beta}$ defined in Section~\ref{sec:2HDM}.}
\( \beta\) and \( \alpha \) respectively describe the inverse time duration of the phase transition and its strength. The former is usually normalized to the Hubble parameter and is given 
by
\begin{equation}
    \frac{\beta}{H_*} =\frac{v_b \eta}{R_*H_*} ,\quad R_* = \left[T_p^3\int_{T_p}^{T_c}dT^\prime\frac{\Gamma(T^\prime)}{{T^\prime}^4 H(T^\prime)}e^{-I(T^\prime)}\right]^{-\frac{1}{3}},
    \label{beta_bubble}
\end{equation}
 where $R_*$ is the mean bubble separation \cite{Turner:1992tz, Enqvist:1991xw} and 
 $\eta = (8\pi)^{1/3}$.
As for the PT strength $\alpha$, it is given by the ratio of the trace anomaly to the radiation energy density~\cite{Espinosa:2010hh}:
\begin{equation}
\label{alpha}
 \alpha=\frac{\epsilon}{\rho_{R}}=\frac{1}{\frac{\pi^2}{30} g_* T_{p}^4}\left(\Delta V-\frac{T_{p}}{4} \Delta s\right)
\end{equation}
where \( \Delta V \) and \( \Delta s \) are the potential and entropy difference between the false and true vacuum at \( T_{p} \), respectively, while \( g_* \) is the number of relativistic degrees of freedom.

\radd{We note that the ratio $v/T$, commonly used in electroweak-baryogenesis studies through the sphaleron-washout criterion $v_n/T_n\gtrsim1$, does not enter our analysis: the CP-conserving 2HDM considered here contains no new source of CP violation, and the gravitational-wave signal is instead controlled by $(\alpha,\beta/H_*)$ evaluated at percolation. The two sets of criteria are correlated but not equivalent, as quantified in the discussion of Figure~\ref{fig:Tc_Tp_vp_Tp} in Section~\ref{sec:phase_diagrams}. Moreover, in the presence of strong supercooling, \(T_n\) may not exist even though \(T_p\) does and the transition ultimately completes~\cite{Athron:2022mmm}.} 
The apparent paradox arises because the two criteria probe different quantities: the nucleation condition counts bubbles and requires their integrated number to reach one per Hubble volume, whereas the percolation condition weights each bubble by the volume it has grown to occupy. The integrated bubble number may therefore remain below unity at all temperatures, leaving \(T_n\) undefined, while the relatively rare bubbles nucleated earlier have sufficient time to grow and overlap, allowing \(I(T)\) to reach the percolation threshold. This does not violate causality, since a Hubble volume is not a causal volume and a causally expanding bubble may become comparable to, or even larger than, the instantaneous Hubble volume. Provided that the physical false-vacuum volume is decreasing, as verified through Eq.~\eqref{eq:volume_decreases} for all points retained in our analysis, the transition can subsequently proceed to completion.

\subsection{Results: phase diagrams of \( \beta/H_* \) and $\alpha$}
\label{sec:phase_diagrams}
\begin{figure}[h!]
    \centering
    \includegraphics[width=0.8\linewidth]{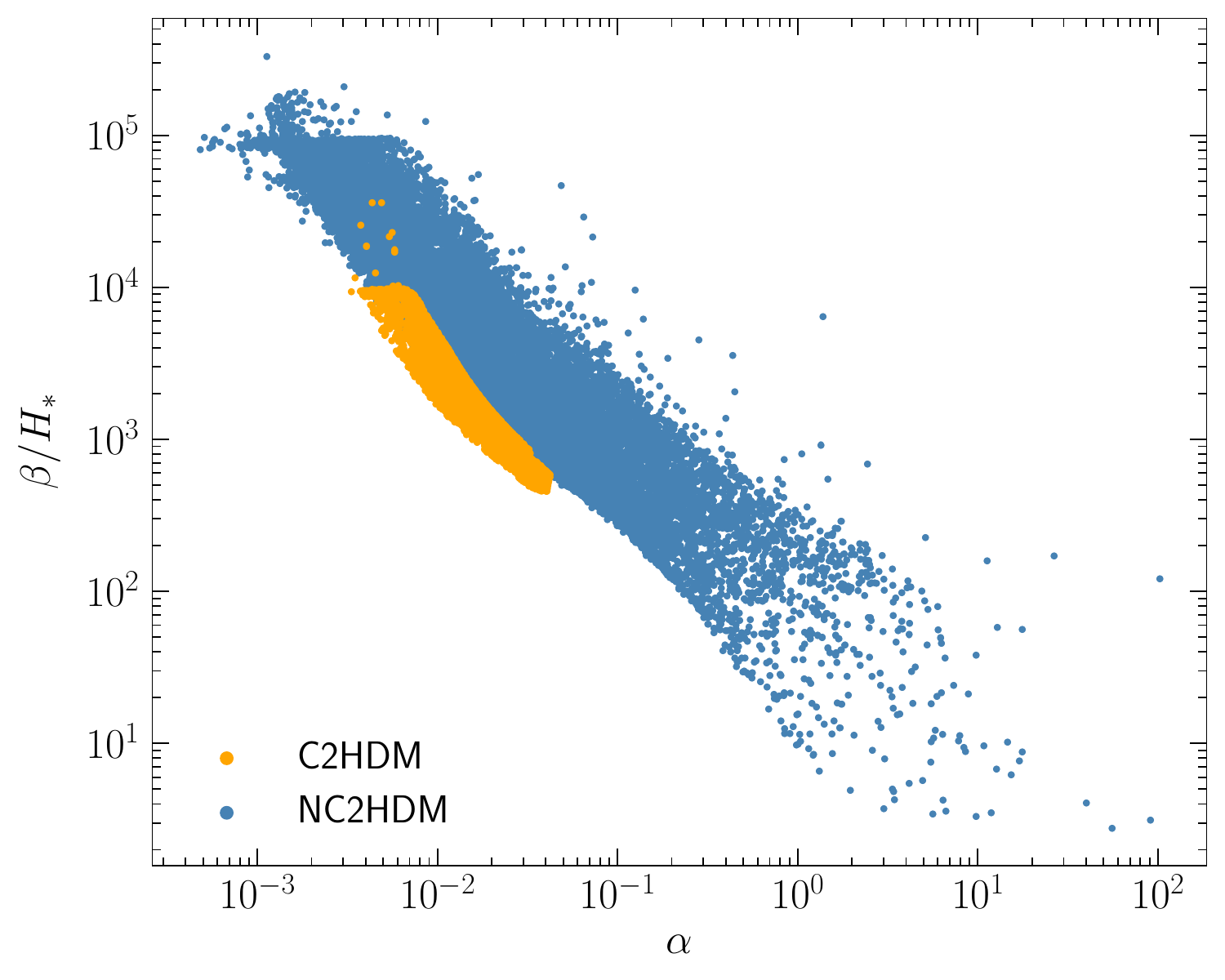}
    \caption{Correlation between the PT strength $\alpha$ and the normalized inverse time PT duration $\beta/H_*$. Blue (orange) points correspond to the NC2HDM (C2HDM).}
    \label{fig:alpha_beta}
\end{figure}

In this section, we present the results of our phase transition analysis in the (N)C2HDM via the phase diagram in Figure~\ref{fig:alpha_beta}, showing the PT strength $\alpha$ on the x-axis and the PT inverse duration $\beta/H_*$ on the y-axis. As expected, we recover the usual anti-correlation between these quantities, that is slower  phase transitions (smaller $\beta/H_*$) are stronger   (larger $\alpha$). Surprisingly, however, the largest value of $\alpha$, or equivalently the lowest value of $\beta/H_*$, is found in the NC2HDM. As shown in the next section, the amount of supercooling in the C2HDM is quite limited, which keeps $\alpha$ comparatively small. \radd{The physical origin of this limitation is structural rather than a matter of parameter counting: once the scalon is identified with the $125$~GeV Higgs boson, the Gildener--Weinberg relation Eq.~(\ref{Mhcon}) fixes the size of the radiative breaking of scale invariance, and hence the depth of the one-loop potential along the flat direction, independently of the remaining scan parameters (see Section~\ref{sec:limited_supercooling} and Appendix~\ref{app:supercooling}).} Figure~\ref{fig:alpha_beta} also shows that the NC2HDM spans a wider region of the $\left(\alpha,\beta/H_*\right)$ than the C2HDM, accommodating both weaker and stronger phase transitions; indeed, the C2HDM domain is nearly contained within the NC2HDM envelope.

\begin{figure}[h!]
    \centering
    \includegraphics[width=\linewidth]{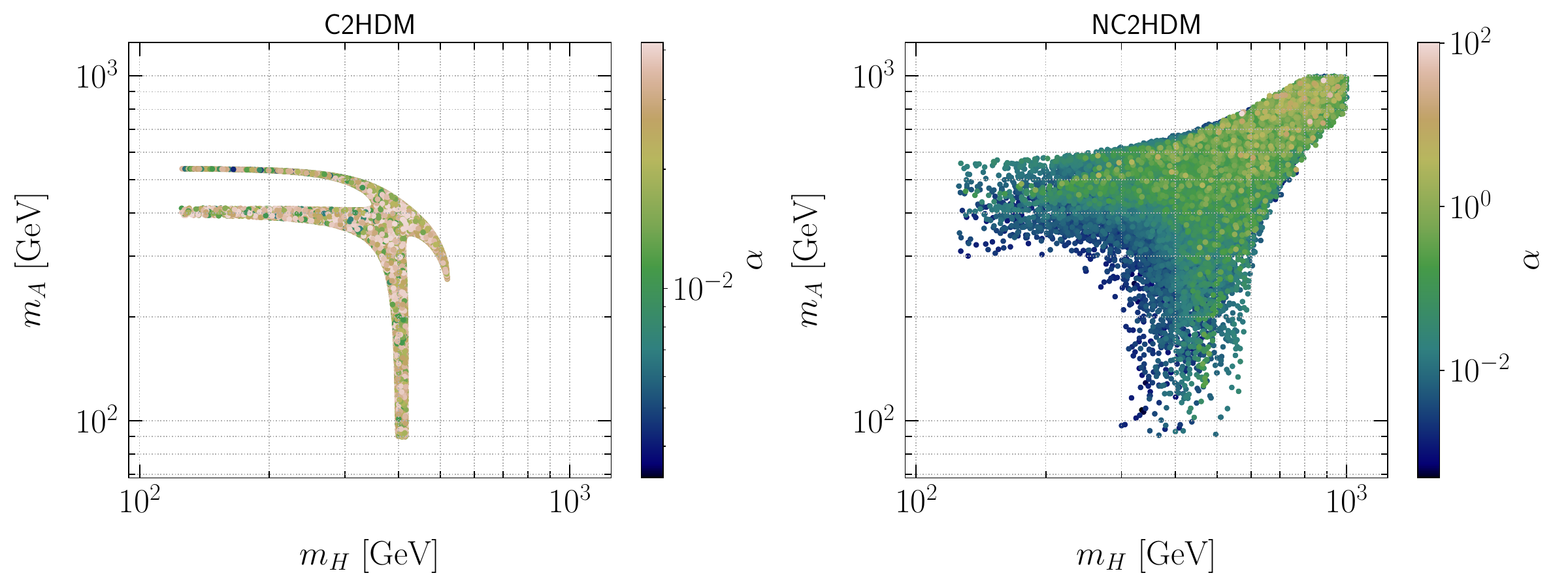}
\includegraphics[width=\linewidth]{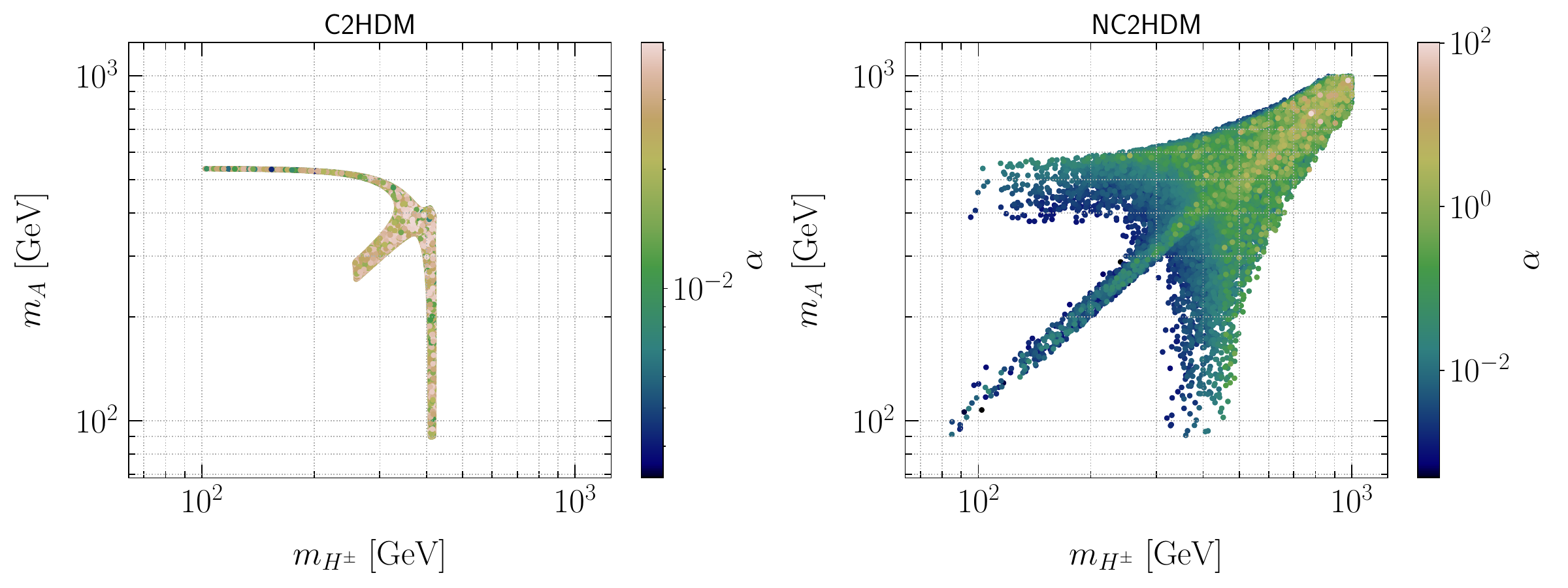}
    \caption{The upper panels show the projection of the parameter space on the plane covered by the masses $m_{H}$ and $m_A$ in the (N)C2HDM, while the lower panels show the projection on the plane $m_{H^\pm}-m_A$. The colour-bar legends show the value of the strength parameter $\alpha$.}
    \label{fig:mH_mHc_mA}
\end{figure}

In the following, we project the parameter space to isolate regions that yield strong first-order phase transitions (FOPTs). Figure~\ref{fig:mH_mHc_mA} shows which values of  physical heavy scalar masses lead to FOPTs. The left column is the allowed parameter space for the C2HDM, while the plots in the right column are for the NC2HDM. Two features emerge. First, the C2HDM projections are nearly contained within those of the NC2HDM.
Second, we observe that in the NC2HDM there is a positive correlation between the heavy scalar mass and the PT strength $\alpha$. Indeed stronger FOPTs tend to be favored for larger scalar masses. Moreover, for large masses, a degenerate mass spectrum yields the strongest FOPTs. By contrast, in the C2HDM the heavy-state masses show little correlation with $\alpha$. In addition, the sum rule in Eq.~(\ref{Mhcon}), together with the requirement $m_H > m_h$,
forbids values of $m_A$ and $m_{H^\pm}$ from becoming  too large (see also Figure~\ref{fig:MH_grid}). Likewise,  this same relation imposes an upper bound on $m_H$. No analogous upper bounds arise in the NC2HDM for $m_H$, $m_A$ and $m_{H^\pm}$. 

Finally, in the NC2HDM the requirement of a FOPT, combined with  
theoretical and electroweak-precision constraints, gives rise to a
pronounced tail of points with $m_A \simeq m_{H^\pm}$ in the
$m_A$--$m_{H^\pm}$ plane, while no analogous tail with $m_A \simeq m_H$
appears in the $m_A$--$m_H$ projection. From a scan perspective, the
origin of this tail can be understood as the intersection of two
selection effects. A first-order transition can be realized over a
rather broad range of scalar masses, so the FOPT condition by itself
does not fix the mass scale very sharply and is relatively insensitive
to moderate splittings among the additional scalars. Before imposing the
FOPT requirement, the combination of theoretical bounds and electroweak
precision data already forces most accepted points to lie close to the
diagonal $m_A \simeq m_{H^\pm}$, i.e.\ to an approximately
custodial-symmetric spectrum, since the $T$ parameter strongly constrains
the mass splitting between these two states. When we then require an 
FOPT, many of these points are removed, in particular those for which
the transition is too weak. Because stronger FOPTs tend to occur at
larger scalar masses, the surviving points are statistically biased
toward the high-mass part of this diagonal band. Nevertheless, the
pre-FOPT distribution is densest along $m_A \simeq m_{H^\pm}$ over the
whole mass range, so even in the low-mass region some points on the
diagonal remain after the cut. The resulting pattern in the
$m_A$--$m_{H^\pm}$ projection therefore appears as an extended tail
along the diagonal.

By contrast, there is no analogous mechanism that
would align the points along the diagonal $m_A \simeq m_H$: this
relation is not protected by a simple approximate symmetry, and $m_H$ is
additionally constrained by the CP-even mass matrix and Higgs signal
strengths. Consequently, parameter points with $m_A \simeq m_H$ that
still realize an FOPT and satisfy all constraints occupy only a much
smaller, localised region in the $m_A$--$m_H$ plane and do not form a
visible tail.


\begin{figure}[h!]
    \centering
    \includegraphics[width=\linewidth]{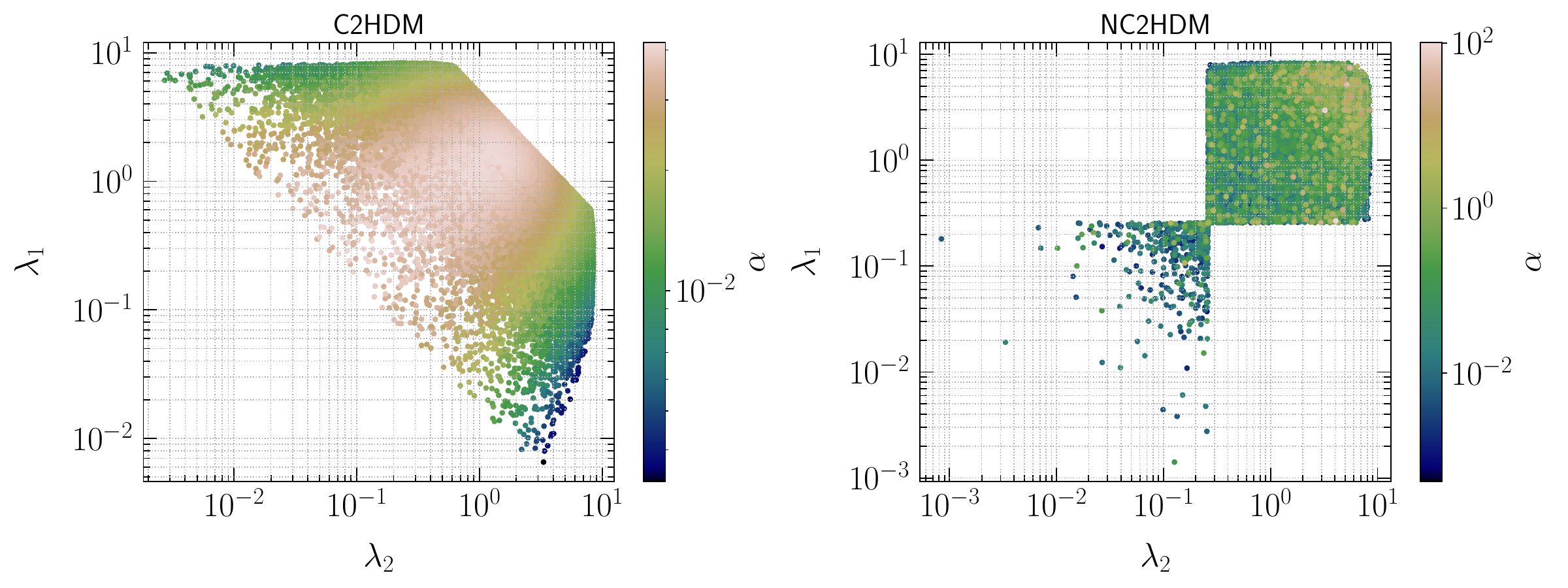}
    \includegraphics[width=\linewidth]{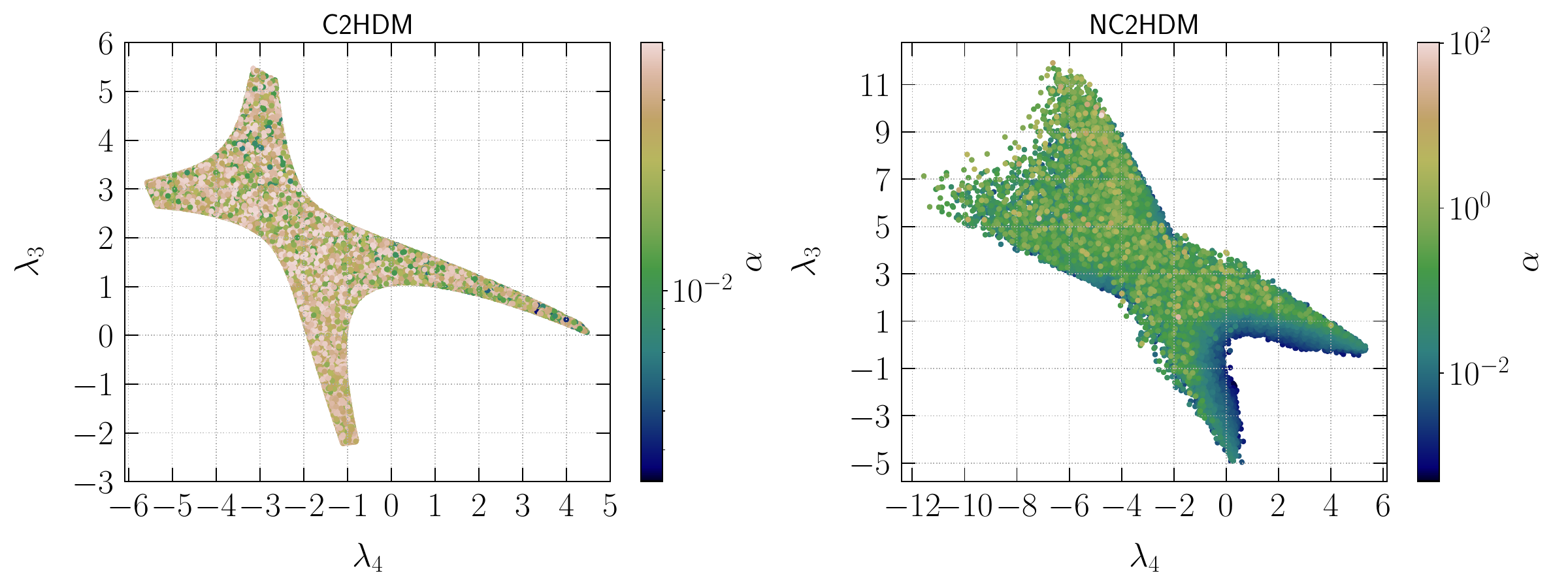}
    \includegraphics[width=\linewidth]{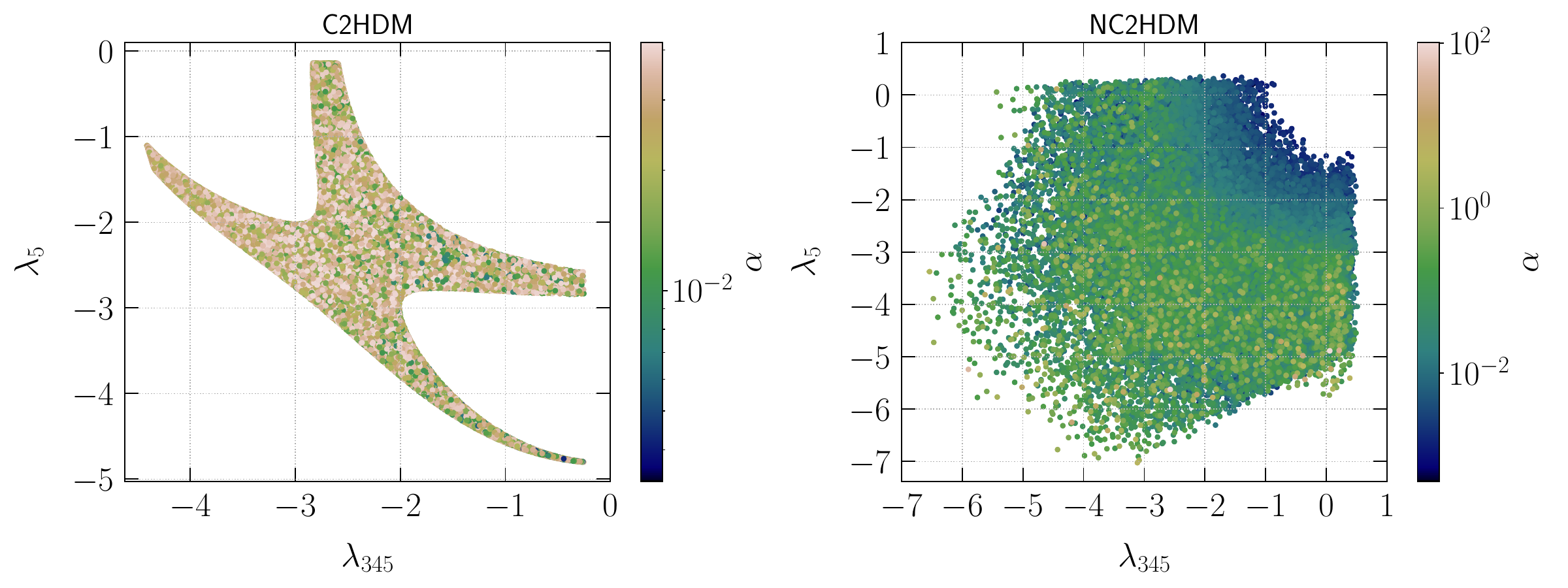}
    \caption{Correlation between quartic couplings in the (N)C2HDM. The upper panels show the correlation between $\lambda_1$ and $\lambda_2$, while the middle panels show the correlation between $\lambda_3$ and $\lambda_4$. Finally the lower panels show the correlation between $\lambda_5$ and $\lambda_{345}$. The colour-bar legends show the value of the strength parameter $\alpha$.}
    \label{fig:lambdas}
\end{figure}

Figure~\ref{fig:lambdas}, shows projections of the viable parameter points onto planes spanned by the quartic couplings in the (N)C2HDM. The first row presents the results for $\lambda_1$ and $\lambda_2$. In the C2HDM, we clearly see a symmetric pattern, in which the strongest FOPTs cluster around the central region (roughly along $\lambda_1\simeq\lambda_2$). Moving away from the middle,  towards larger values of $\lambda_1$ or $\lambda_2$, yields a smaller $\alpha$. This indicates that the strongest FOPTs are obtained for moderate values of the mixing angle: $\tan\tilde{\beta}\sim 1$. 
Indeed from the parametrisation Eq.~\eqref{flatmin}, increasing $\tan\tilde{\beta}$ increases $\lambda_1$ and decreases $\lambda_2$, i.e. it moves points towards the upper-left of the $(\lambda_1,\lambda_2)$ plane where $\alpha$ is smaller; decreasing $\tan\tilde{\beta}$ moves points towards the lower-right, again yielding weaker transitions.
In the NC2HDM, the correlation is less pronounced: the $(\lambda_1,\lambda_2)$ plane exhibits an approximately horizontal boundary at $\lambda_1\simeq 0.25$ and a vertical boundary at $\lambda_2\simeq 0.25$, reached in the limits $\tan\tilde{\beta}\to 0$ and $\tan\tilde{\beta}\to\infty$, respectively.
Indeed, in these two limits, we can compute $\lambda_{1,2}\rightarrow m_h^2/v^2\simeq 0.25$, via Eq.~\eqref{mcoup2}. 
Intermediate values of $\tan\tilde{\beta}$ are found within the quadrants made by these two boundaries. Finally, we can observe that the lower left quadrant is less populated due to the stability condition Eq.~\eqref{eq:bfb} and stronger FOPTs are found for simultaneously large $\lambda_1$ and $\lambda_2$. In the second row, there is no apparent correlation between $\lambda_{3,4}$ and $\alpha$ in the (N)C2HDM and the covered parameter spaces in both models have a similar shape, although the range of values is larger in the NC2HDM. 
Finally, in the third row the two models differ: while $\alpha$ shows little dependence on $\lambda_5$ in the C2HDM, in the NC2HDM the weakest transitions lie near the origin of the $(\lambda_5,\lambda_{345})$ plane, i.e. for small $|\lambda_5|$ and $|\lambda_{345}|$.

\begin{figure}[h!]
    \centering
    \includegraphics[width=\linewidth]{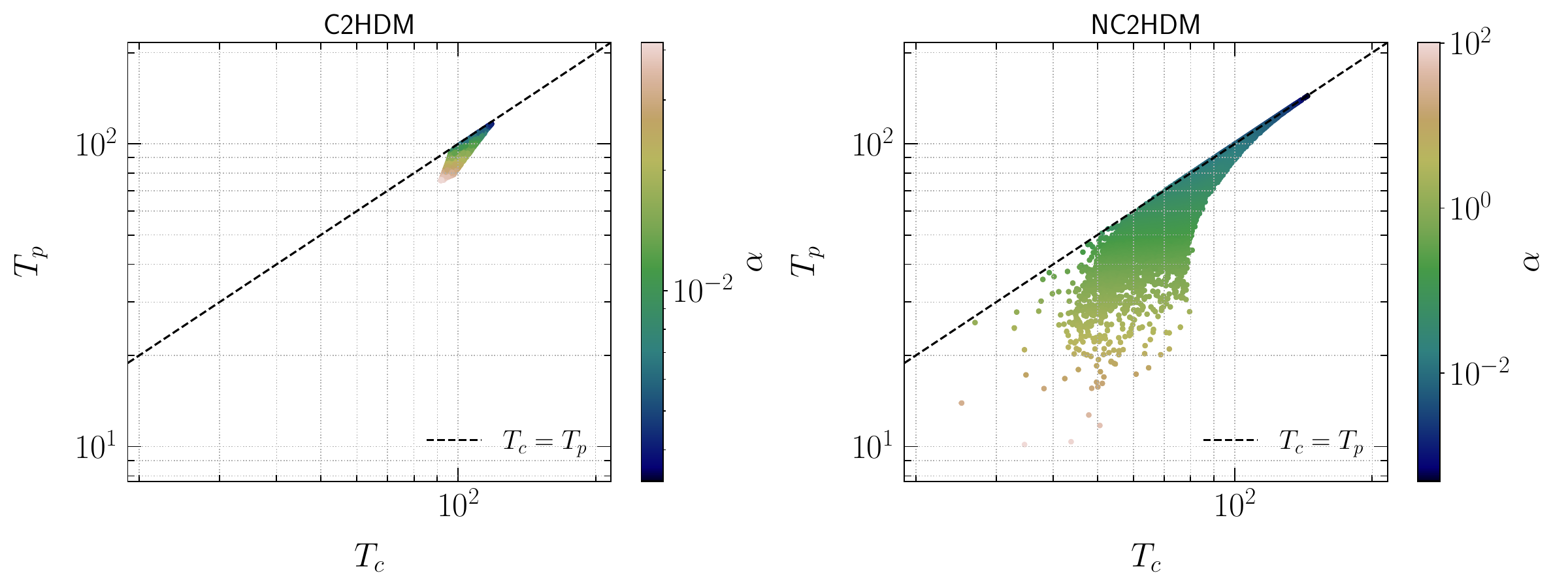}
    \includegraphics[width=\linewidth]{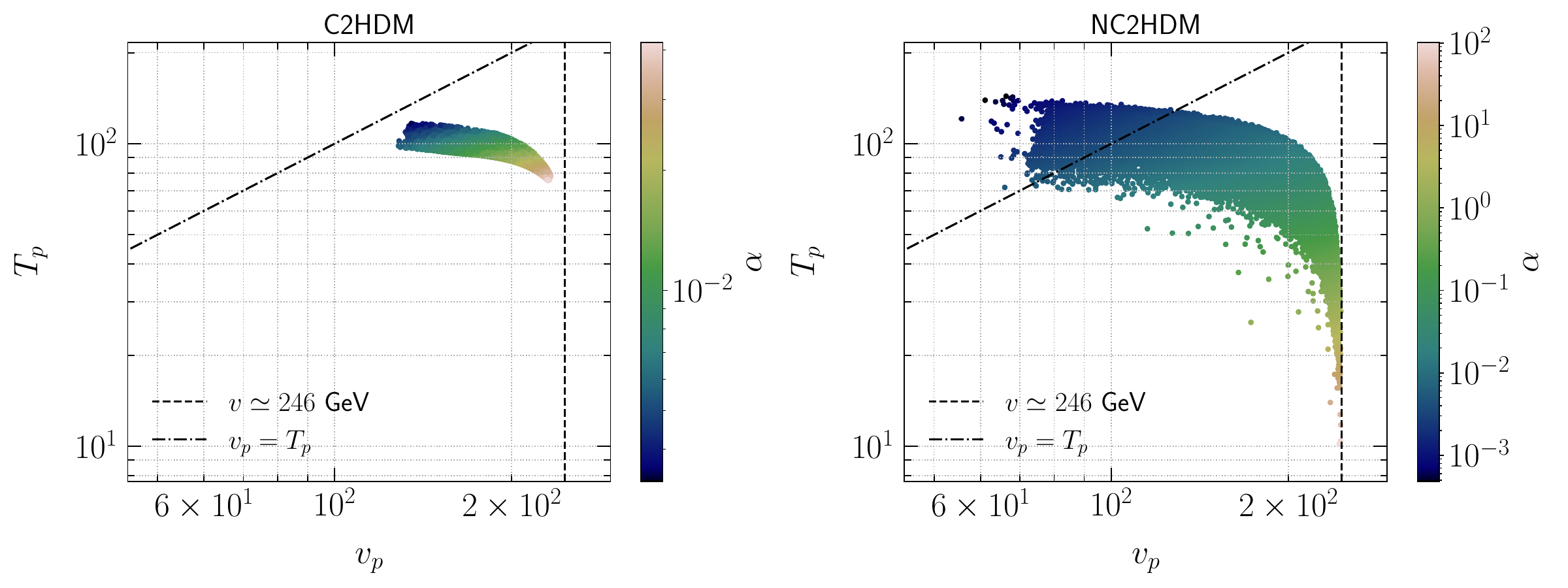}
    \caption{The upper panels show the correlation between the critical temperature $T_c$ and the percolation temperature $T_p$ in the (N)C2HDM, while the lower panels show the correlation between $T_p$ and $v_p$, the position of the global minimum in the effective potential at $T=T_p$. The colour-bar legends show the value of the strength parameter $\alpha$. \radd{The dash-dotted line in the lower panels denotes $v_p=T_p$, making the conventional strong-transition criterion $v_p/T_p>1$ visually explicit; note that a sizable fraction of the points with $v_p/T_p>1$ nevertheless features $\alpha\lesssim10^{-2}$ (see main text).}}
    \label{fig:Tc_Tp_vp_Tp}
\end{figure}

The amount of supercooling in both the conformal and non-conformal 2HDM is depicted in Figure~\ref{fig:Tc_Tp_vp_Tp}. In the first row, the two crucial temperatures for the PT analysis,
$T_p$ and $T_c$, 
 are presented. The diagonal dashed line corresponds to $T_p=T_c$, which physically means that, as soon as the false vacuum stops being the global minimum,   tunneling through the barrier occurs. In such a case the PT would be extremely weak. Points that are far below this dashed lines naturally lead to a stronger PT, as indicated by the larger value of $\alpha$. Indeed, $T_p\ll T_c$ indicates substantial supercooling: the universe remains trapped for a long time in the metastable false vacuum even though the true vacuum is already energetically favored. The longer it remains in this metastable phase, the larger the amount of supercooling. A long supercooling enhances the released vacuum energy and, at the same time, reduces the radiation energy density at the transition, so that $\alpha$ is correspondingly larger. 
In the second row in Figure~\ref{fig:Tc_Tp_vp_Tp}, $T_c$ is now replaced by $v(T=T_p)\equiv v_p$, the VEV at the true vacuum at $T=T_p$ and where the vertical dashed line denotes the value of this VEV at $T=0$, $v(T=0)\equiv v\simeq 246$ GeV. \radd{The dash-dotted line denotes $v_p=T_p$ and makes the conventional strong-transition (sphaleron-washout) criterion $v_p/T_p>1$ visually explicit; it can be contrasted directly with the colour-coded strength parameter $\alpha$, which measures the released energy normalised to the radiation bath.} We observe that larger $v_p$ tends to correlate with larger $\alpha$. Indeed, as the universe cools, the position of the true vacuum shifts to larger field values, approaching its zero-temperature value $v$. At the same time, the true vacuum becomes progressively deeper, increasing $\Delta V$. Therefore, larger $v_p$ typically corresponds to a stronger phase transition. Note also that as $v_p$ approaches $v$, the percolation temperature $T_p$ drops rapidly, enhancing the amount of supercooling.

\radd{The $v_p=T_p$ reference line also exposes the inequivalence between the conventional order-parameter criterion $v/T\gtrsim1$ and the energy-based measure $\alpha$. In both panels, a large fraction of the points satisfying $v_p/T_p>1$ features only $\alpha=\mathcal{O}(10^{-3}\text{--}10^{-2})$: in the NC2HDM, for instance, about $44\%$ ($12\%$) of the points with $v_p/T_p>1.5$ have $\alpha<0.02$ ($\alpha<0.01$), and $58\%$ of those with $v_p/T_p>2$ still have $\alpha<0.05$. The two criteria indeed probe different quantities. The order-parameter ratio, which controls the sphaleron energy in the broken phase, can be of order one already at $T\simeq T_c$, where the free-energy difference between the two phases vanishes identically; by contrast, $\alpha\simeq30\,\Delta V(T_p)/(\pi^2g_*T_p^4)$ (neglecting the entropy contribution in Eq.~\eqref{alpha}) vanishes in the limit $T_p\to T_c$ of weak supercooling and is further diluted by the large number of relativistic degrees of freedom, $g_*\sim10^2$. In the C2HDM the released energy is additionally loop-suppressed, $\Delta V=\mathcal{O}(\mathcal{B}\,v_p^4)$ with $\mathcal{B}\simeq m_h^2/(8v^2)\simeq0.03$, so that $\alpha=\mathcal{O}(10^{-3})$ for $v_p/T_p\simeq1.5$. Conversely, since $\Delta V\lesssim\bar{\lambda}\,v_p^4/4$ with an effective depth coefficient $\bar{\lambda}\lesssim\mathcal{O}(1)$, a sizable strength enforces $v_p/T_p\gtrsim\left[4\pi^2g_*\,\alpha/(30\bar{\lambda})\right]^{1/4}$, e.g.\ $v_p/T_p\gtrsim1.6$ for $\alpha=0.05$ and $\bar{\lambda}=1$: all strong transitions lie deep in the $v_p/T_p>1$ region, while the converse does not hold. Within our scan, $v_p/T_p>1$ therefore behaves as a necessary but not sufficient condition for a transition that is strong in the gravitational-wave sense, and we accordingly characterise the strength by $\alpha$ (together with $\beta/H_*$) evaluated at percolation.}

\subsection{Supercooling in C2HDM: role of the scalon mass}
\label{sec:limited_supercooling}

For the parameter space under consideration, we identify the CP-even scalar i.e.~scalon (the pseudo–Goldstone boson of broken scale invariance) with the observed SM-like Higgs boson and thus fix \( m_h \simeq 125\,\mathrm{GeV} \). This requirement provides a significant constraint in the C2HDM scan. In the Gildener--Weinberg setup, $m_h$ is generated radiatively and is well approximated by Eq.~\eqref{Mhcon}, which correlates the heavy-scalar masses and restricts the viable input ranges.
In practice, for each fixed choice of $m_h$ (set to $125~\mathrm{GeV}$ in the SM-like case, or varied over the grid values in the relaxed analysis), we scan over $(m_A,\,m_{H^\pm},\,\tan\tilde{\beta})$ and retain only points that satisfy Eq.~\eqref{Mhcon} (with $m_H>m_h$) together with all constraints in Sec.~\ref{sec:constraints}.

\begin{figure}[t!]
    \centering
    \includegraphics[width=\linewidth]{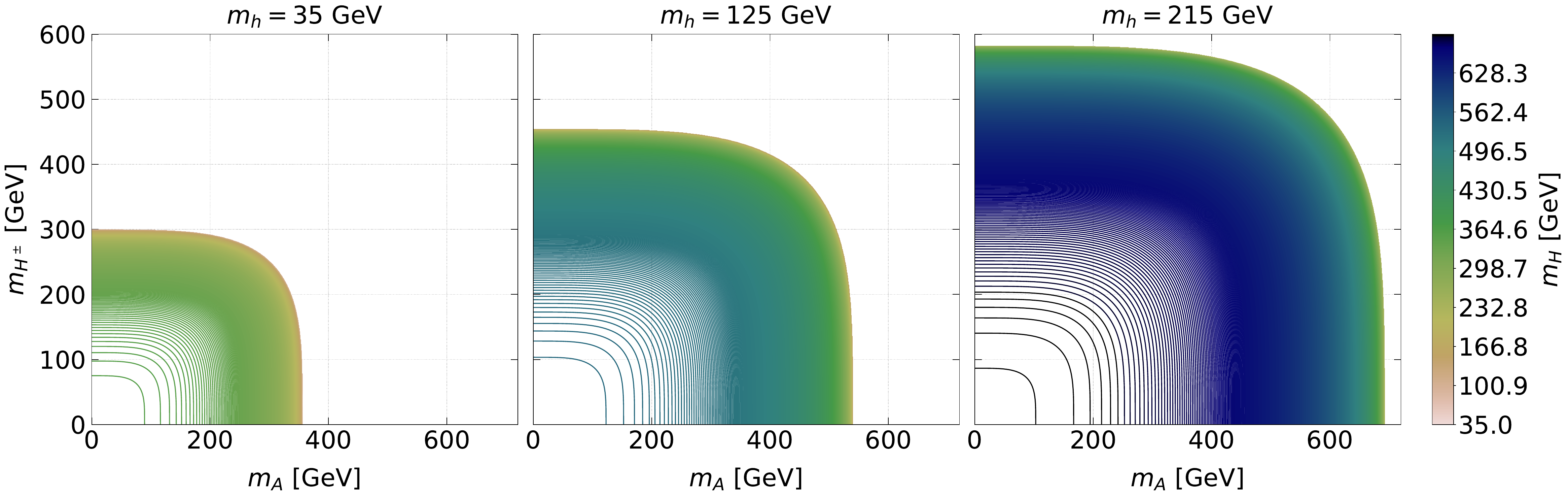}
    \caption{Impact of the scalon mass $m_h$ on the heavy scalar masses $m_H$ (indicated by the color bar), $m_A$ and $m_{H^\pm}$ in the C2HDM. The left, middle and right panels show the results for $m_h=35$ GeV, $m_h=125$ GeV, $m_h=215$ GeV, respectively. We can see that the smaller the scalon mass, the smaller the upper bounds on $m_H$, $m_A$ and $m_{H^\pm}$.}
    \label{fig:MH_grid}
\end{figure}


\begin{table}[h!]
\centering
\renewcommand{\arraystretch}{1.2}
\small
\begin{tabular}{c|cccc}
\hline
$m_h$ [GeV] & $T_c$ [GeV] & $T_p$ [GeV] & $\alpha$ & $1 - T_p/T_c$ \\
\hline
35  & [43.38---59.92]   & [5.55---28.46]     & [0.246---134.941]   & [0.482---0.876] \\
65  & [61.39---71.52]   & [32.48---54.58]    & [0.054---0.401]     & [0.226---0.475] \\
95  & [78.99---90.22]   & [57.11---79.38]    & [0.021---0.083]     & [0.112---0.278] \\
125 & [96.12---108.52]  & [78.36---102.10]   & [0.011---0.037]     & [0.057---0.186] \\
155 & [113.00---124.35] & [97.69---120.05]   & [0.007---0.021]     & [0.033---0.136] \\
185 & [129.62---138.40] & [116.16---135.53]  & [0.005---0.014]     & [0.019---0.105] \\
215 & [145.40---150.99] & [133.66---148.31]  & [0.004---0.010]     & [0.011---0.085] \\
\hline
\end{tabular}
\caption{Results are shown for the intersection set of benchmark parameter sets specified by $(m_H,\,m_A,\,m_{H^\pm})$. Among the 200 parameter sets scanned for each fixed $m_h$, only 69 remain viable across all $m_h$ benchmarks considered here. The amount of supercooling is quantified by $1 - T_p/T_c$.}
\label{mhcolum}
\end{table}

A common expectation is that classically scale-invariant potentials generically lead to a prolonged supercooling stage. \radd{This expectation~\cite{Witten:1980ez,Konstandin:2011dr,Jinno:2016knw,Marzola:2017jzl,Iso:2017rjs} rests on two features of scale-invariant potentials: since the tree-level potential contains no negative mass term, the origin is stabilised by the thermal masses at any $T>0$ and destabilised only by the loop-generated logarithmic terms, so the barrier persists down to arbitrarily low temperatures and no spinodal temperature exists at which the transition would be forced to complete; moreover, along the radiatively lifted flat direction the tunneling action decreases only logarithmically as the temperature drops, so that nucleation is generically postponed far below $T_c$, deep into the vacuum-dominated regime where $\alpha\gg1$; a detailed discussion of both features is provided in Appendix~\ref{app:supercooling}.} Our results show that this is not automatic: the amount of supercooling is controlled by how weakly scale invariance is broken by quantum effects, which in the present model is conveniently parameterised by the scalon mass $m_h$. \radd{Quantitatively, the Gildener--Weinberg relation Eq.~(\ref{Mhcon}) gives $\mathcal{B}\simeq m_h^2/(8v^2)\simeq0.03$ for $m_h=125$~GeV: for such a sizeable radiative lifting the flat direction is in fact far from flat, the tunneling action drops rapidly below $T_c$, and the supercooling mechanism described above operates only marginally ($1-T_p/T_c\lesssim0.19$, Table~\ref{mhcolum}). In this sense the observed Higgs boson is too heavy for the conformal supercooling mechanism --- which explains why the C2HDM populates the weaker-transition region of the $(\alpha,\beta/H_*)$ plane in Figure~\ref{fig:alpha_beta} --- while the mechanism progressively re-emerges as $m_h$ is lowered.} To make this dependence explicit, we repeat the scan for fixed values
$m_h\in\{35,\allowbreak 65,\allowbreak 95,\allowbreak 125,\allowbreak 155,\allowbreak 185,\allowbreak 215\}\,\mathrm{GeV}$,
selecting 200 benchmark parameter sets for each choice. The resulting allowed ranges of $(m_H,\,m_A,\,m_{H^\pm})$ as $m_h$ is varied are displayed in Fig.~\ref{fig:MH_grid}.

\begin{figure}[h!]
    \centering
    \includegraphics[width=0.8\linewidth
    ]
    {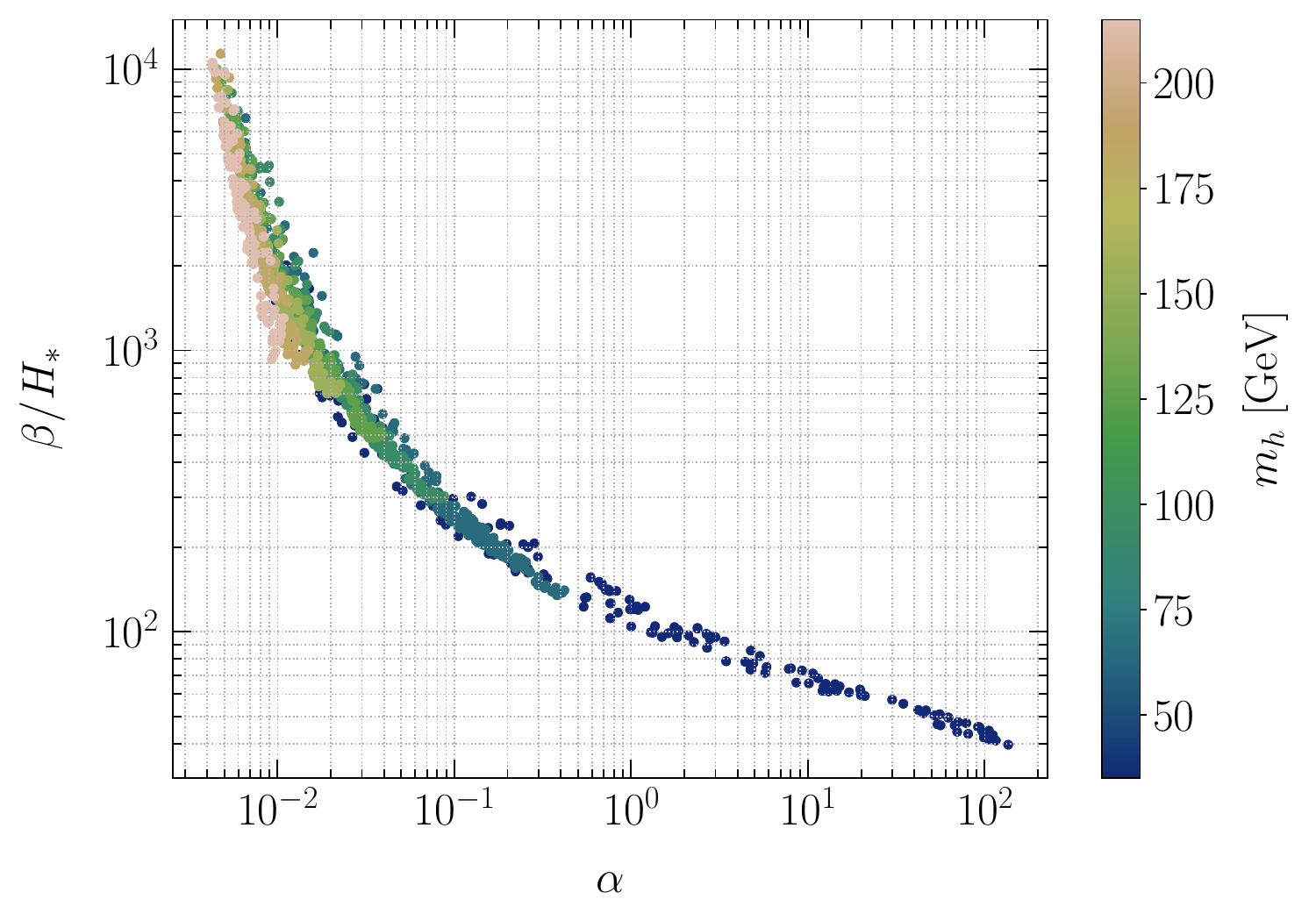}
    \caption{Correlation between the PT strength $\alpha$ and $\beta/H_*$ in the C2HDM for points from Table~\ref{mhcolum}. The color bar indicates different values of the scalon mass, considered as a free parameter in this analysis.}
    \label{diffscalon}
\end{figure}


The trend is clear (see Fig.~\ref{diffscalon}): when the scalon is significantly lighter than
$125~\mathrm{GeV}$ (e.g.\ $m_h \simeq 35~\mathrm{GeV}$), the phase transition typically exhibits much stronger supercooling, whereas for $m_h \gtrsim 200~\mathrm{GeV}$ the FOPT becomes substantially weaker. Importantly, a benchmark parameter set in $(m_H,\,m_A,\,m_{H^\pm})$ that is viable for one $m_h$ may be excluded for another. For a consistent comparison across different $m_h$, we therefore focus on the intersection set, identifying 69 benchmark parameter sets that remain viable for all scanned $m_h$ values. The corresponding phase-transition parameter ranges are summarised in Table~\ref{mhcolum}: for each fixed $m_h$, we quote the resulting intervals of $T_c$, $T_p$,
$\alpha$, and the supercooling measure $1 - T_p/T_c$, which makes explicit the overall tendency that smaller $m_h$ typically leads to stronger supercooling. \radd{This tendency parallels the well-known Standard-Model result that a lighter Higgs yields a stronger transition, the order parameter scaling as $v_c/T_c\simeq2E/\lambda_T\propto1/m_h^2$ (with $E$ the coefficient of the thermally generated cubic term and $\lambda_T$ the quartic) --- one of the facts preventing electroweak baryogenesis in the SM for the measured Higgs mass. The parallel is, however, only qualitative, because the underlying mechanisms, and their scalings with $m_h$, differ. In the SM the barrier is the thermal cubic $-E\,T\phi^3$ and $m_h$ enters through the ratio $E/\lambda_T$, a power law, with the transition completing promptly below $T_c$ so that $\alpha$ stays modest. In the C2HDM, instead, $m_h^2\simeq8\mathcal{B}v^2$ fixes the coefficient $\mathcal{B}$ of the entire radiatively generated potential (Section~\ref{sec:C2HDM-CW}), which controls both the persistence of the barrier and the logarithmic descent of the tunneling action (Appendix~\ref{app:supercooling}); since completion is governed by the exponentially small rate $\Gamma\propto e^{-S_3/T}$ with $S_3/T\propto1/\mathcal{B}$ up to logarithms, the percolation temperature --- and hence $\alpha\propto\Delta V/T_p^4$ --- responds \emph{exponentially}, rather than power-like, to the scalon mass, as made fully explicit in Eq.~\eqref{eq:app_alphaB}. Table~\ref{mhcolum} bears this out: lowering $m_h$ by a factor $\simeq6$ (from $215$ to $35$~GeV) would enhance the strength by only a factor $\simeq40$ under the SM $1/m_h^2$ law, whereas $\alpha$ actually grows by about four orders of magnitude, from $\mathcal{O}(10^{-2})$ to $\mathcal{O}(10^2)$. The steep scalon-mass dependence of the conformal 2HDM is therefore the imprint of supercooling, not of the SM thermal-barrier ratio $E/\lambda_T$.}

To summarise, supercooling in the C2HDM is not an automatic consequence of classical scale invariance: it is strongly enhanced only in the near-conformal regime where the one-loop breaking is sufficiently mild, which in our setup is conveniently tracked by a light scalon mass $m_h$. 
In the next section, we use the percolation-based quantities evaluated at $T_p$ (in particular $\alpha$ and $\beta/H_*$) to compute the resulting stochastic gravitational-wave spectra and assess their detectability.

\section{Gravitational-wave predictions}
\label{sec:gravitational_waves}
\subsection{GW template and mapping from phase-transition parameters}

The total contribution to the stochastic gravitational-wave background consists of the contribution from bubble collision, overlapping of sound waves and magnetohydrodynamic (MHD) turbulence. The resulting GW power spectrum from a first-order phase transition in the early universe is thus, at least approximately, a linear superposition of the three aforementioned contributions: 
\begin{equation}
h^2\Omega_{\text{GW},0}\simeq h^2\Omega_{\text{col},0} + h^2\Omega_{\text{sw},0} + h^2\Omega_{\text{turb},0}.
\end{equation}
\radd{The bubble-collision term is controlled by the fraction $\kappa_\phi$ of the released vacuum energy stored in the gradient energy of the walls, its amplitude scaling as $h^2\Omega_{\text{col},0}\propto[\kappa_\phi\,\alpha/(1+\alpha)]^2\,(H_*/\beta)^2$~\cite{Alanne:2019bsm}. A sizable $\kappa_\phi\simeq1-\alpha_\infty/\alpha$ requires the driving pressure to overcome the leading-order plasma friction, i.e.\ $\alpha>\alpha_\infty$, with $\alpha_\infty$ the maximal strength that this friction can balance~\cite{Bodeker:2009qy,Alanne:2019bsm}; since $\alpha_\infty\propto(v_p/T_p)^2$ while $\alpha\propto(v_p/T_p)^4$, this occurs only for strongly supercooled transitions. For all the points considered here we find $\alpha<\alpha_\infty$, up to a single marginal exception whose collision peak remains at the percent level of the plasma-sourced one, so that $\kappa_\phi\simeq0$: the released energy is transferred to the plasma, the signal is dominated by sound waves and MHD turbulence (computed with $v_b=1$ as a terminal wall velocity), and we neglect $h^2\Omega_{\text{col},0}$ in the numerical spectra of Fig.~\ref{fig:GW}.}




We use the stochastic GW background template for first-order phase transitions with moderate supercooling~\cite{Caprini:2024hue}:



\begin{equation}
    h^2\Omega_\text{GW}^\text{DBPL} = h^2\Omega_2\frac{S(f)}{S(f_2)},\quad S(f)=N\left(\frac{f}{f_1}\right)^{n_1}\left[1+\left(\frac{f}{f_1}\right)^{a_1}\right]^\frac{n_2-n_1}{a_1}\left[1+\left(\frac{f}{f_2}\right)^{a_2}\right]^\frac{n_3-n_2}{a_2},   
    \label{eq:templateII}
\end{equation}
where $S(f)$ is a dimensionless double-broken power-law (DBPL) shape function with turnover frequencies $f_1$ and $f_2$.
It interpolates between three asymptotic power laws,
$S(f)\propto f^{n_1}$ for $f\ll f_1$,
$S(f)\propto f^{n_2}$ for $f_1\ll f\ll f_2$,
and $S(f)\propto f^{n_3}$ for $f\gg f_2$,
while $a_1$ and $a_2$ control the smoothness of the turnovers around $f_1$ and $f_2$.
The ratio $S(f)/S(f_2)$ fixes the normalization such that $h^2\Omega_\text{GW}^\text{DBPL}(f_2)=h^2\Omega_2$.
The constant $N$ provides an overall normalization of $S(f)$; however, it cancels in the ratio $S(f)/S(f_2)$,
so that the spectrum is, by construction, normalized to $h^2\Omega_\text{GW}^\text{DBPL}(f_2)=h^2\Omega_2$
(independently of the choice of $N$).

The phase-transition dynamics enters the GW template through the two parameters $\alpha$ and $\beta/H_*$
introduced in Sec.~\ref{intro_beta}. The strength parameter $\alpha$ controls the fraction of released energy that is
converted into bulk kinetic energy of the plasma via 
\begin{equation}\label{Keq}
    K\simeq 0.6\, \frac{\kappa  \alpha}{(1+\alpha)}
\end{equation}
The inverse duration $\beta/H_*$ is encoded through the mean bubble separation $R_*$: using Eq.~\eqref{beta_bubble},
\begin{equation}
    \frac{1}{H_*R_*}=\frac{\beta/H_*}{v_b\,\eta}\,,
\end{equation}
so that the break frequencies in Eq.~\eqref{eq:templateII} scale as $f_{1,2}\propto (H_*R_*)^{-1}\propto \beta/H_*$,
whereas the amplitudes scale as
$h^2\Omega\propto K^2(H_*R_*)$.
Consequently, larger $\alpha$ and smaller $\beta/H_*$ lead to stronger GW signals and lower peak frequencies,
consistent with the trends observed in the $(\alpha,\beta/H_*)$ phase diagrams.

For the sound-wave contribution, one has $n_1=3,n_2=1,n_3=-3,a_1=2,a_2=4$ and
\begin{equation}
    f_1\simeq 0.2\frac{ H_{*,0}}{H_*R_*},\quad f_2\simeq 0.5\frac{ H_{*,0}}{\Delta_b H_*R_*},\quad \Delta_b = \frac{\vert v_b-c_s\vert}{\max(v_b,c_s)},
\end{equation}
where $v_b$ is the bubble-wall speed and $c_s$ is the speed of sound in the plasma. The parameter $\Delta_b$ measures the relative velocity difference between the
bubble wall and the sound front and therefore sets the characteristic thickness of the acoustic
shell, $\Delta R \sim \Delta_b R_*$. Consequently, the first break $f_1$ is associated with the
mean bubble separation scale $R_*$, while the second break $f_2$ is set by the shell thickness.
For a relativistic plasma, $c_s^2=1/3$, so that $\Delta_b=1-1/\sqrt{3}\simeq 0.42$ for $v_b=1$.
The redshift factor for the GW amplitude is given by $h^2F_\text{GW,0}\simeq 1.64\times 10^{-5}(100/g_*)^{1/3}$ and the Hubble parameter at the time of GW production redshifted to today is given by
\begin{equation}
    H_{*,0} \simeq 1.65\times 10^{-5}\text{~Hz~}\left(\frac{g_*}{100}\right)^{1/6}\left(\frac{T_*}{100\text{~GeV}}\right).
\end{equation}

The GW amplitude at $f_2$ is 
\begin{equation}
    h^2\Omega_2=\frac{1}{\pi}\left(\sqrt{2}+\frac{2f_2/f_1}{1+f_2^2/f_1^2}\right)h^2F_{\text{GW},0}A_\text{sw}K^2(H_*R_*)\Upsilon,
\end{equation}
with $A_{\rm sw}\simeq 0.11$ a dimensionless calibration constant that fixes the overall
normalization of the sound-wave GW spectrum in the DBPL template (as recommended in
Ref.~\cite{Caprini:2024hue}),
and $K$, 
the fraction of released energy converted into
bulk kinetic energy of the plasma (with $\kappa(\alpha)$ taken from Ref.~\cite{Espinosa:2010hh}),
is given by \eqref{Keq}.
The suppression factor $\Upsilon$ is given in a radiation-dominated universe by~\cite{Guo:2020grp}
\begin{equation}
\Upsilon=1-\frac{1}{\sqrt{2\tau_\text{sh}H_* + 1}},\quad\tau_\text{sh}\equiv \frac{R_*}{\bar{U}_f},
\end{equation}
which is a generalisation of $\Upsilon = \min(H_*\tau_\text{sh},1)$ used in~\cite{Caprini:2024hue}. The enthalpy-weighted root-mean-square of the fluid velocity is $\bar{U}_f=\sqrt{K/\Gamma}$, with the mean adiabatic index $\Gamma=4/3$ for a relativistic fluid. 

Regarding the MHD contribution to the stochastic GW background power spectrum, the spectral shape in Eq.~(\ref{eq:templateII}) is obtained using $n_1=3,n_2=1,n_3=-8/3,a_1=4,a_2\simeq 2.15$, and the following frequency breaks:
\begin{equation}
    f_1=\frac{\bar v_A}{\mathcal{N}}\frac{H_{*,0}}{H_*R_*},\quad f_2  \simeq 2.2\frac{ H_{*,0}}{H_*R_*}, \quad \bar v_A = \sqrt{\epsilon\frac{K}{\Gamma}},
\end{equation}
with $\mathcal{N}\simeq 2$ and with $\epsilon$ the fraction of overall kinetic energy in bulk motion that is
converted to MHD turbulence; we consider $\epsilon =0.5$. The amplitude at $f_2$ is given by
\begin{equation}
    h^2\Omega_2 = h^2 F_\text{GW,0}A_\text{MHD}(\epsilon K)^2(H_*R_*)^2,\quad A_\text{MHD}=3\times 2.2\frac{\mathcal{A}}{4\pi^2}\times2^{-11/3a_2},
\end{equation}
with $\mathcal{A}\simeq0.085$.

\subsection{Conformal vs non-conformal GW signals and detectability}

\begin{figure}[t!]
    \centering
    \includegraphics[width=0.7\linewidth]{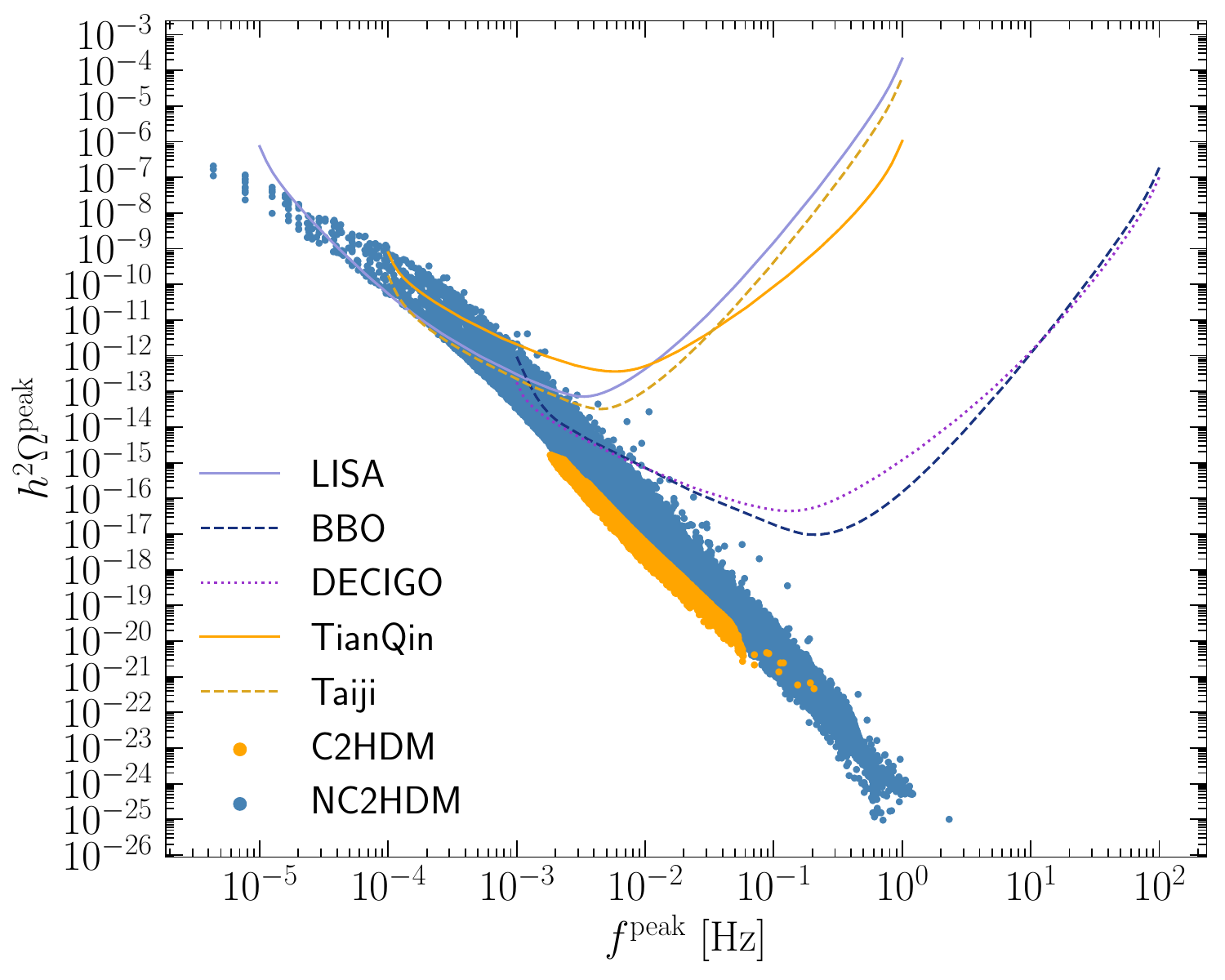}
    \caption{Peak of the GW power spectrum $h^2\Omega_\text{GW}^\text{peak}$ as a function of its associated peak frequency $f^\text{peak}$. Blue (orange) points correspond to the NC2HDM (C2HDM). Sensitivity curves of the space-based GW detectors LISA, BBO, DECIGO, TianQin and Taiji are also shown.}
    \label{fig:GW}
\end{figure}

In Fig.~\ref{fig:GW} we plot the peak amplitude of the stochastic gravitational-wave spectrum from first-order phase transitions, $h^2\Omega_\text{GW}^\text{peak}$, together with the corresponding
peak frequency $f^\text{peak}$. The overall behaviour closely follows that in Fig.~\ref{fig:alpha_beta}, since the peak amplitude is enhanced for larger $\alpha$ and smaller $\beta/H_*$, whereas the characteristic frequency scales approximately as $f^\text{peak}\propto \beta/H_*$.  We see that  stronger PTs shift $\beta/H_*$ to the upper-left direction, while simultaneously shifting $f^\text{peak}$ to lower peak frequencies. 

In addition, we also present the power-law-integrated sensitivity curves~\cite{Thrane:2013oya} of LISA~\cite{2017arXiv170200786A}, DECIGO~\cite{Kawamura:2011zz}, BBO~\cite{Corbin:2005ny}, TianQin~\cite{TianQin:2015yph} and Taiji~\cite{Hu:2017mde} assuming an observation time of 4 years and signal-to-noise ratio of 10. We find that only the NC2HDM provides benchmark points that give sufficiently strong electroweak first-order phase transitions to reach the sensitivity of near-future space-based gravitational-wave detectors, such as LISA. By contrast, the C2HDM typically produces weaker signals
due to its limited supercooling, placing its predicted GW spectra below the LISA band. Nevertheless, one can note that, in terms of $h^2\Omega_\text{GW}^\text{peak}$, it could be probed by detectors with higher sensitivity, such as DECIGO or BBO; however, their optimal sensitivity lies at higher frequencies than the typical $f^\text{peak}$ predicted in the C2HDM. \radd{We have also verified that the present constraint from the LVK O1--O4a stochastic-background search and the big-bang-nucleosynthesis (BBN) bound on the total gravitational-wave energy density are both too weak to constrain the parameter space considered here, and we therefore do not display them in Fig.~\ref{fig:GW}. The LVK band probes frequencies well above the peak frequencies obtained in our scan and, since the predicted peak amplitude decreases with increasing peak frequency, the points closest to it carry the weakest signals --- as expected for electroweak-scale transitions, whose signals peak in the mHz band. The BBN bound, an integral constraint on the total gravitational-wave energy density~\cite{Maggiore:1999vm,Kahniashvili:2021kls}, would affect only the very upper part of the $(f^\text{peak},h^2\Omega_\text{GW}^\text{peak})$ plane and is satisfied by all our spectra.}

\section{Conclusion}
\label{sec:conclusion}

We considered the CP-conserving two-Higgs-doublet model in two realizations: a classically conformal setup and a non-conformal setup with explicit tree-level quadratic mass terms.
After constructing the finite-temperature effective potential in the (N)C2HDM, we performed parameter scans as in Section~\ref{sec:scan}, imposing the theoretical and experimental constraints from Sections~\ref{sec:theor_const} and~\ref{sec:exp_const}. 
For each viable point, we analysed the electroweak phase-transition dynamics along the one-field trajectory and extracted the thermodynamic
scales $T_c$ and $T_p$. In particular, in the presence of supercooling we evaluated the phase-transition strength $\alpha$ and inverse duration $\beta/H_*$ at the percolation temperature $T_p$, which provides
a robust input for   subsequent gravitational-wave predictions.

As shown in Figure~\ref{fig:alpha_beta}, we find that the strongest first-order phase transitions (larger $\alpha$ and smaller $\beta/H_*$) occur in the NC2HDM, while the C2HDM typically exhibits a
more limited amount of supercooling. This challenges the common expectation that imposing classical scale invariance at tree level generically implies deep supercooling. 
In the C2HDM, identifying the scalon with the observed SM-like Higgs fixes $m_h\simeq 125~\mathrm{GeV}$ and therefore fixes the size of
the radiative breaking of scale invariance in the Gildener--Weinberg setup. In Section~\ref{sec:limited_supercooling}, we relax this identification by treating $m_h$ as an external input and show (Figure~\ref{diffscalon}) that supercooling is strongly enhanced only when the radiative breaking is sufficiently mild, i.e.~for a
light scalon. Thus, classical conformal symmetry alone does not guarantee strong supercooling; rather, the amount of supercooling is controlled by how weakly scale invariance is broken radiatively (in our setup, tracked by the scalon mass).

Finally, in Figure~\ref{fig:GW} we computed the stochastic gravitational-wave spectra associated with electroweak first-order phase transitions and compared them with the projected sensitivities of future
space-based interferometers. We find that only the NC2HDM yields benchmark points that can enter the sensitivity region of near-future detectors such as LISA (and also TianQin and Taiji). By contrast, the
C2HDM signal typically lies below the LISA band due to its limited supercooling; nevertheless, in terms of peak amplitude it could be accessible to more sensitive missions such as DECIGO or BBO, although the
corresponding peak frequencies tend to lie below their most sensitive bands.

\section*{Acknowledgments}

ZWW, JWL and NB are partially supported by the National Natural Science Foundation of China (Grant No. 12475105).
TGS and RBM are  supported by the Natural Sciences \& Engineering Research Council of Canada (NSERC Grant SAPIN-2021-00024).

\appendix

{
\section{Classical scale invariance and strong supercooling}
\label{app:supercooling}

In this appendix we collect, in a self-contained way, the two features of classically scale-invariant potentials that underlie the common expectation of strong supercooling, and we explain why the effect is truncated in the C2HDM once the scalon is identified with the observed Higgs boson. Throughout, a generic polynomial (non-conformal) potential is contrasted with a radiatively generated (conformal) one along the relevant transition direction.

\subsection{Persistence of the barrier}
\label{app:barrier}

For a polynomial potential, the standard high-temperature form along the order-parameter direction reads (see e.g.~\cite{quiros1999finite})
\begin{equation}
V_{\rm poly}(\phi,T)\;\simeq\;D\left(T^2-T_0^2\right)\phi^2\;-\;E\,T\,\phi^3\;+\;\frac{\lambda_T}{4}\,\phi^4 ,
\label{eq:app_Vpoly}
\end{equation}
with $D,E>0$ and $\lambda_T$ slowly varying functions of the couplings, and $T_0$ the \emph{spinodal} temperature, at which the curvature at the origin, $M^2(T)=2D\,(T^2-T_0^2)$, vanishes. For $T<T_0$ the origin turns into a maximum: the barrier has disappeared and the field rolls over classically. A first-order transition must therefore complete within the finite window $T_0<T<T_c$, and the amount of supercooling is bounded by construction.

In a classically conformal potential the dimension-two term is absent. The curvature at the origin is $+c\,T^2$ at any $T>0$, with $c>0$ the thermal-mass coefficient of Eq.~\eqref{var}, while the destabilisation of the symmetric phase relative to the true vacuum is driven only by the loop-generated logarithmic terms. The origin therefore remains a local minimum at all temperatures: no spinodal temperature exists, and the false vacuum can, in principle, survive down to $T\to0$.

\subsection{Temperature dependence of the tunneling action}
\label{app:action}

Close to $T_c$ the two cases behave identically: the minima are nearly degenerate, the bounce is in the thin-wall regime, and
\begin{equation}
\frac{S_3}{T}\;\simeq\;\frac{16\pi}{3}\,\frac{\sigma^3}{T\,\Delta V(T)^2},
\qquad
\Delta V(T)\simeq L\,\frac{T_c-T}{T_c},
\label{eq:app_thinwall}
\end{equation}
with $\sigma$ the wall tension and $L$ the latent heat, so that $S_3/T\to\infty$ as $T\to T_c$. The difference lies in how the action decreases once the Universe cools below $T_c$.

For the polynomial potential, rescaling $\phi=T\chi$ gives
\begin{equation}
\frac{V_{\rm poly}}{T^4}\;=\;D\left(1-\frac{T_0^2}{T^2}\right)\chi^2\;-\;E\,\chi^3\;+\;\frac{\lambda_T}{4}\,\chi^4 :
\label{eq:app_poly_resc}
\end{equation}
the rescaled shape depends on the temperature through the \emph{power-law} combination $T_0^2/T^2$, which moreover changes sign at $T=T_0$. Correspondingly, the bounce action takes the form~\cite{Dine:1992wr,quiros1999finite}
\begin{equation}
\frac{S_3}{T}\;=\;\frac{M^3(T)}{E^2\,T^3}\;F(\kappa),
\qquad
\kappa\;\equiv\;\frac{\lambda_T\,M^2(T)}{2E^2T^2},
\label{eq:app_S3poly}
\end{equation}
where $F$ is a known, numerically fitted function that diverges in the thin-wall limit $\kappa\to1$ (i.e.\ $T\to T_c$) and stays finite as $\kappa\to0$. As the temperature decreases from $T_c$ to $T_0$, $\kappa$ decreases monotonically from $1$ to $0$ and $S_3/T$ falls \emph{steeply} from infinity to zero, $S_3/T\propto M^3(T)\to0$ at the spinodal. The nucleation condition $S_3/T\sim\mathcal{O}(10^2)$ is thus met soon after $T_c$ and, in any case, the transition cannot be delayed past $T_0$: supercooling is doubly limited, by the steep descent and by the spinodal floor.

For the conformal potential the same rescaling yields, schematically,
\begin{equation}
\frac{V_{\rm conf}}{T^4}\;\simeq\;\frac{c}{2}\,\chi^2\;-\;E\,\chi^3\;+\;\mathcal{B}\,\chi^4\left[\ln\frac{T^2\chi^2}{v^2}-\frac{1}{2}\right],
\label{eq:app_conf_resc}
\end{equation}
where the quadratic and cubic terms are of purely thermal origin and the quartic-logarithmic term descends from the Gildener--Weinberg potential. The explicit temperature dependence of the rescaled shape now resides \emph{only} in $\ln(T/v)$: by dimensional analysis, in a scale-invariant theory $V/T^4$ can depend on the temperature only through the logarithmically running couplings. Consequently $S_3/T$ is nearly a plateau, decreasing only logarithmically as the Universe cools. If this plateau sits above the nucleation threshold at $T\lesssim T_c$, the transition stalls: nucleation and percolation are postponed by many $e$-folds of cooling (with the $O(4)$ bounce eventually competing with the thermal one, cf.\ Eq.~\eqref{eq:min_prescription}), while the released vacuum energy approaches its zero-temperature value, $\Delta V\to\mathcal{B}v^4/2$, and the radiation bath dilutes as $T^4$. The strength parameter then grows as $\alpha\simeq\Delta V/\rho_R\propto(v/T_p)^4$, which is the origin of the very large values of $\alpha$ attainable in supercooled conformal models~\cite{Witten:1980ez,Konstandin:2011dr,Iso:2017rjs}.

\subsection{Truncation of the mechanism in the C2HDM}
\label{app:truncation}

Both the height of the plateau and the rate of its logarithmic descent are governed by the single scale-breaking coefficient $\mathcal{B}$: it sets the depth of the radiative minimum, $|V_{\rm conf}(v)|=\mathcal{B}v^4/2$, as well as the drift per $e$-fold of cooling of the rescaled shape in Eq.~\eqref{eq:app_conf_resc}. In the C2HDM, $\mathcal{B}$ is not a free parameter: the Gildener--Weinberg relation Eq.~\eqref{Mhcon} ties it to the scalon mass, $\mathcal{B}\simeq m_h^2/(8v^2)$. For $m_h=125$~GeV this gives $\mathcal{B}\simeq0.03$ --- a sizeable radiative lifting, for which the ``flat'' direction is in fact far from flat: the logarithmic descent is fast, the nucleation condition is met after only a moderate amount of cooling, and the supercooling stage is truncated ($1-T_p/T_c\lesssim0.19$, Table~\ref{mhcolum}). For a light scalon the mechanism is instead fully operative: for $m_h=35$~GeV one has $\mathcal{B}\simeq2.5\times10^{-3}$, and we indeed find $\alpha$ up to $\mathcal{O}(10^2)$ with $T_p$ as low as $\sim0.1\,T_c$ (Section~\ref{sec:limited_supercooling}). In this sense, the observed Higgs boson is too heavy for the conformal supercooling mechanism in the 2HDM.

This dependence can be made fully explicit. In the vacuum-dominated regime the released energy approaches the depth of the radiative minimum, $\Delta V\to\mathcal{B}v^4/2$ (for the Gildener--Weinberg potential $V=\mathcal{B}\phi^4[\ln(\phi^2/v^2)-1/2]$, whose curvature at the minimum consistently reproduces $m_h^2=8\mathcal{B}v^2$), so that the strength parameter of Eq.~\eqref{alpha} reduces to
\begin{equation}
\alpha\;\simeq\;\frac{30\,\Delta V(T_p)}{\pi^2g_*T_p^4}\;\simeq\;\frac{15\,\mathcal{B}}{\pi^2g_*}\left(\frac{v}{T_p}\right)^{4}.
\label{eq:app_alphaTp}
\end{equation}
The supercooling depth $v/T_p$ is itself fixed by $\mathcal{B}$ through the completion condition $S_3(T_p)/T_p \simeq c_*=\mathcal{O}(10^2)$ (cf.\ the nucleation estimate $S_3/T\simeq140$ of Section~\ref{sec:PT_dynamics}). Deep in the supercooled regime, tunneling proceeds through the thermal barrier of Eq.~\eqref{eq:app_conf_resc} with the negative, logarithmically running quartic $\lambda_{\rm eff}(T)\simeq8\mathcal{B}\ln(v/T)$; using the thick-wall bounce action $S_3\simeq18.9\,m_{\rm th}/\lambda_{\rm eff}$~\cite{Linde:1980tt,Linde:1981zj}, with $m_{\rm th}\simeq\sqrt{c}\,T$ the thermal mass at the origin, the completion condition yields
\begin{equation}
\ln\frac{v}{T_p}\;\simeq\;\frac{18.9\,\sqrt{c}}{8\,c_*\,\mathcal{B}}\;\equiv\;\frac{\kappa}{\mathcal{B}}
\qquad\Longleftrightarrow\qquad
T_p\;\simeq\;v\,e^{-\kappa/\mathcal{B}}\,,
\label{eq:app_TpB}
\end{equation}
and, combined with Eq.~\eqref{eq:app_alphaTp}, the explicit exponential law
\begin{equation}
\alpha(\mathcal{B})\;\simeq\;\frac{15\,\mathcal{B}}{\pi^2g_*}\;e^{4\kappa/\mathcal{B}}\,,
\qquad
\frac{{\rm d}\ln\alpha}{{\rm d}\ln\mathcal{B}}\;=\;1-\frac{4\kappa}{\mathcal{B}}\,,
\label{eq:app_alphaB}
\end{equation}
which is large and negative for $\mathcal{B}\ll4\kappa$: a heavier scalon suppresses the transition strength exponentially, not as a power. Equations~\eqref{eq:app_TpB} and~\eqref{eq:app_alphaB} hold at fixed thermal coefficients; along the physical C2HDM grid the heavy-scalar masses --- and with them the thermal-mass coefficient $c$ and the thermally induced cubic --- grow together with $\mathcal{B}$ through the Gildener--Weinberg relation Eq.~\eqref{Mhcon}, which softens the effective inverse power of $\mathcal{B}$ in the exponent; empirically, the strongest-supercooling points of Table~\ref{mhcolum} are well described by $\ln(v/T_p)\simeq0.2/\sqrt{\mathcal{B}}$. The exponential character is nevertheless manifest: as $\mathcal{B}$ decreases from $9.5\times10^{-2}$ ($m_h=215$~GeV) to $2.5\times10^{-3}$ ($m_h=35$~GeV), $\ln(v/T_p)$ grows from $\simeq0.6$ to $\simeq3.8$ and $\alpha$ spans four orders of magnitude; at the same time, the master relation Eq.~\eqref{eq:app_alphaTp} reproduces the largest $\alpha$ at each $m_h$ in Table~\ref{mhcolum} to within a factor of $1.6$ (for $g_*\simeq107$), with a mild overestimate that shrinks as the supercooling deepens --- precisely the behaviour expected from the entropy contribution to $\epsilon$, neglected in Eq.~\eqref{eq:app_alphaTp}, which is largest for weak supercooling.

It is instructive to contrast this scalon-mass dependence with the Standard Model, where a lighter Higgs also strengthens the transition, but through the thermally induced cubic barrier of Eq.~\eqref{eq:app_Vpoly}: there the order parameter obeys the \emph{power law} $v_c/T_c\simeq2E/\lambda_T\propto1/m_h^2$ and, since the barrier disappears at the spinodal temperature $T_0$, the transition completes promptly below $T_c$, keeping $\alpha$ modest. In the conformal case the scalon mass instead sets the single coefficient $\mathcal{B}$, and hence --- through $S_3/T\propto1/\mathcal{B}$ (up to logarithms) and the exponential nucleation rate $\Gamma\propto e^{-S_3/T}$ --- the percolation temperature and the amount of supercooling. The transition strength $\alpha$ therefore responds \emph{exponentially}, not power-like, to $m_h$ (Eq.~\eqref{eq:app_alphaB}): this is the origin of the four-orders-of-magnitude variation of $\alpha$ across the grid of Table~\ref{mhcolum} for only a factor $\simeq6$ in $m_h$ (Section~\ref{sec:limited_supercooling}), to be compared with the factor $\simeq40$ that the SM $1/m_h^2$ law would give.
}

\bibliographystyle{JHEP}
\bibliography{2hdm_v2}

@article{ATLAS:2012yve,
    author = "Aad, Georges and others",
    collaboration = "ATLAS",
    title = "{Observation of a new particle in the search for the Standard Model Higgs boson with the ATLAS detector at the LHC}",
    eprint = "1207.7214",
    archivePrefix = "arXiv",
    primaryClass = "hep-ex",
    reportNumber = "CERN-PH-EP-2012-218",
    doi = "10.1016/j.physletb.2012.08.020",
    journal = "Phys. Lett. B",
    volume = "716",
    pages = "1--29",
    year = "2012"
}

@article{CMS:2012qbp,
    author = "Chatrchyan, Serguei and others",
    collaboration = "CMS",
    title = "{Observation of a New Boson at a Mass of 125 GeV with the CMS Experiment at the LHC}",
    eprint = "1207.7235",
    archivePrefix = "arXiv",
    primaryClass = "hep-ex",
    reportNumber = "CMS-HIG-12-028, CERN-PH-EP-2012-220",
    doi = "10.1016/j.physletb.2012.08.021",
    journal = "Phys. Lett. B",
    volume = "716",
    pages = "30--61",
    year = "2012"
}

@article{Kajantie:1996mn,
    author = "Kajantie, K. and Laine, M. and Rummukainen, K. and Shaposhnikov, Mikhail E.",
    title = "{Is there a~ hot electroweak phase transition at $m_H \gtrsim m_W$?}",
    eprint = "hep-ph/9605288",
    archivePrefix = "arXiv",
    reportNumber = "CERN-TH-96-126, HD-THEP-96-15, IUHET-333",
    doi = "10.1103/PhysRevLett.77.2887",
    journal = "Phys. Rev. Lett.",
    volume = "77",
    pages = "2887--2890",
    year = "1996"
}

@article{Witten:1984rs,
    author = "Witten, Edward",
    title = "{Cosmic Separation of Phases}",
    reportNumber = "PRINT-84-0400 (IAS,PRINCETON)",
    doi = "10.1103/PhysRevD.30.272",
    journal = "Phys. Rev. D",
    volume = "30",
    pages = "272--285",
    year = "1984"
}

@article{Hogan:1986dsh,
    author = "Hogan, C. J.",
    title = "{Gravitational radiation from cosmological phase transitions}",
    doi = "10.1093/mnras/218.4.629",
    journal = "Mon. Not. Roy. Astron. Soc.",
    volume = "218",
    number = "4",
    pages = "629--636",
    year = "1986"
}

@article{LIGOScientific:2016aoc,
    author = "Abbott, B. P. and others",
    collaboration = "LIGO Scientific, Virgo",
    title = "{Observation of Gravitational Waves from a Binary Black Hole Merger}",
    eprint = "1602.03837",
    archivePrefix = "arXiv",
    primaryClass = "gr-qc",
    reportNumber = "LIGO-P150914",
    doi = "10.1103/PhysRevLett.116.061102",
    journal = "Phys. Rev. Lett.",
    volume = "116",
    number = "6",
    pages = "061102",
    year = "2016"
}

@article{Basler_2017,
	doi = {10.1007/jhep02(2017)121},
  
	url = {https://doi.org/10.1007%2Fjhep02%282017%29121},
  
	year = 2017,
	month = {feb},
  
	publisher = {Springer Science and Business Media {LLC}
},
  
	volume = {2017},
  
	number = {2},
  
	author = {P. Basler and M. Krause and M. Mühlleitner and J. Wittbrodt and A. Wlotzka},
  
	title = {Strong first order electroweak phase transition in the {CP}-conserving 2HDM revisited},
  
	journal = {Journal of High Energy Physics}
}

@article{Lee_2012,
	doi = {10.1103/physrevd.86.035004},
  
	url = {https://doi.org/10.1103%2Fphysrevd.86.035004},
  
	year = 2012,
	month = {aug},
  
	publisher = {American Physical Society ({APS})},
  
	volume = {86},
  
	number = {3},
  
	author = {Jae Sik Lee and Apostolos Pilaftsis},
  
	title = {Radiative corrections to scalar masses and mixing in a scale invariant two Higgs doublet model},
  
	journal = {Physical Review D}
}

@article{Huang_2020,
	doi = {10.1103/physrevd.102.095025},
  
	url = {https://doi.org/10.1103%2Fphysrevd.102.095025},
  
	year = 2020,
	month = {nov},
  
	publisher = {American Physical Society ({APS})},
  
	volume = {102},
  
	number = {9},
  
	author = {W.{\hspace{0.167em}
}C. Huang and F. Sannino and Z.{\hspace{0.167em}}W. Wang},
  
	title = {Gravitational waves from Pati-Salam dynamics},
  
	journal = {Physical Review D}
}

@article{Deshpande:1977rw,
    author = "Deshpande, Nilendra G. and Ma, Ernest",
    title = "{Pattern of Symmetry Breaking with Two Higgs Doublets}",
    reportNumber = "OITS-81",
    doi = "10.1103/PhysRevD.18.2574",
    journal = "Phys. Rev. D",
    volume = "18",
    pages = "2574",
    year = "1978"
}

@misc{quiros1999finite,
      title={Finite temperature field theory and phase transitions}, 
      author={Mariano Quiros},
      year={1999},
      eprint={hep-ph/9901312},
      archivePrefix={arXiv},
      primaryClass={hep-ph}
}

@article{Zhang_2021,
	doi = {10.1007/jhep05(2021)160},
  
	url = {https://doi.org/10.1007%2Fjhep05%282021%29160},
  
	year = 2021,
	month = {may},
  
	publisher = {Springer Science and Business Media {LLC}
},
  
	volume = {2021},
  
	number = {5},
  
	author = {Zhao Zhang and Chengfeng Cai and Xue-Min Jiang and Yi-Lei Tang and Zhao-Huan Yu and Hong-Hao Zhang},
  
	title = {Phase transition gravitational waves from pseudo-Nambu-Goldstone dark matter and two Higgs doublets},
  
	journal = {Journal of High Energy Physics}
}

@article{Kobakhidze_2017,
	doi = {10.1140/epjc/s10052-017-5132-y},
  
	url = {https://doi.org/10.1140%2Fepjc%2Fs10052-017-5132-y},
  
	year = 2017,
	month = {aug},
  
	publisher = {Springer Science and Business Media {LLC}
},
  
	volume = {77},
  
	number = {8},
  
	author = {Archil Kobakhidze and Cyril Lagger and Adrian Manning and Jason Yue},
  
	title = {Gravitational waves from a supercooled electroweak phase transition and their detection with pulsar timing arrays},
  
	journal = {The European Physical Journal C}
}

@article{Wainwright_2012,
	doi = {10.1016/j.cpc.2012.04.004},
  
	url = {https://doi.org/10.1016%2Fj.cpc.2012.04.004},
  
	year = 2012,
	month = {sep},
  
	publisher = {Elsevier {BV}
},
  
	volume = {183},
  
	number = {9},
  
	pages = {2006--2013},
  
	author = {Carroll L. Wainwright},
  
	title = {{CosmoTransitions}: Computing cosmological phase transition temperatures and bubble profiles with multiple fields},
  
	journal = {Computer Physics Communications}
}

@article{PhysRevD.46.2384,
  title = {Bubble nucleation in first-order inflation and other cosmological phase transitions},
  author = {Turner, Michael S. and Weinberg, Erick J. and Widrow, Lawrence M.},
  journal = {Phys. Rev. D},
  volume = {46},
  issue = {6},
  pages = {2384--2403},
  numpages = {0},
  year = {1992},
  month = {Sep},
  publisher = {American Physical Society},
  doi = {10.1103/PhysRevD.46.2384},
  url = {https://link.aps.org/doi/10.1103/PhysRevD.46.2384}
}

@article{M_gevand_2017,
	doi = {10.1016/j.nuclphysb.2017.03.009},
  
	url = {https://doi.org/10.1016%2Fj.nuclphysb.2017.03.009},
  
	year = 2017,
	month = {jun},
  
	publisher = {Elsevier {BV}
},
  
	volume = {919},
  
	pages = {74--109},
  
	author = {Ariel M{\'{e}}gevand and Santiago Ram{\'{\i}}rez},
  
	title = {Bubble nucleation and growth in very strong cosmological phase transitions},
  
	journal = {Nuclear Physics B}
}

@article{Cai_2017,
	doi = {10.1088/1475-7516/2017/08/004},
  
	url = {https://doi.org/10.1088%2F1475-7516%2F2017%2F08%2F004},
  
	year = 2017,
	month = {aug},
  
	publisher = {{IOP} Publishing},
  
	volume = {2017},
  
	number = {08},
  
	pages = {004--004},
  
	author = {Rong-Gen Cai and Misao Sasaki and Shao-Jiang Wang},
  
	title = {The gravitational waves from the first-order phase transition with a dimension-six operator},
  
	journal = {Journal of Cosmology and Astroparticle Physics}
}

@article{Ellis_2019,
	doi = {10.1088/1475-7516/2019/04/003},
  
	url = {https://doi.org/10.1088%2F1475-7516%2F2019%2F04%2F003},
  
	year = 2019,
	month = {apr},
  
	publisher = {{IOP} Publishing},
  
	volume = {2019},
  
	number = {04},
  
	pages = {003--003},
  
	author = {John Ellis and Marek Lewicki and Jos{\'{e}
} Miguel No},
  
	title = {On the maximal strength of a first-order electroweak phase transition and its gravitational wave signal},
  
	journal = {Journal of Cosmology and Astroparticle Physics}
}

@article{TianQin:2015yph,
    author = "Luo, Jun and others",
    collaboration = "TianQin",
    title = "{TianQin: a space-borne gravitational wave detector}",
    eprint = "1512.02076",
    archivePrefix = "arXiv",
    primaryClass = "astro-ph.IM",
    doi = "10.1088/0264-9381/33/3/035010",
    journal = "Class. Quant. Grav.",
    volume = "33",
    number = "3",
    pages = "035010",
    year = "2016"
}

@article{Linde:1980tt,
    author = "Linde, Andrei D.",
    title = "{Fate of the False Vacuum at Finite Temperature: Theory and Applications}",
    reportNumber = "LEBEDEV-80-92",
    doi = "10.1016/0370-2693(81)90281-1",
    journal = "Phys. Lett. B",
    volume = "100",
    pages = "37--40",
    year = "1981"
}

@article{Linde:1981zj,
    author = "Linde, Andrei D.",
    title = "{Decay of the False Vacuum at Finite Temperature}",
    reportNumber = "LEBEDEV-81-265",
    doi = "10.1016/0550-3213(83)90072-X",
    journal = "Nucl. Phys. B",
    volume = "216",
    pages = "421",
    year = "1983",
    note = "[Erratum: Nucl.Phys.B 223, 544 (1983)]"
}

@article{Branco_2012,
	doi = {10.1016/j.physrep.2012.02.002},
  
	url = {https://doi.org/10.1016%2Fj.physrep.2012.02.002},
  
	year = 2012,
	month = {jul},
  
	publisher = {Elsevier {BV}
},
  
	volume = {516},
  
	number = {1-2},
  
	pages = {1--102},
  
	author = {G.C. Branco and P.M. Ferreira and L. Lavoura and M.N. Rebelo and Marc Sher and Jo{\~{a}}o P. Silva},
  
	title = {Theory and phenomenology of two-Higgs-doublet models},
  
	journal = {Physics Reports}
}

@article{Cline:1996mga,
    author = "Cline, James M. and Lemieux, Pierre-Anthony",
    title = "{Electroweak phase transition in two Higgs doublet models}",
    eprint = "hep-ph/9609240",
    archivePrefix = "arXiv",
    reportNumber = "MCGILL-96-16",
    doi = "10.1103/PhysRevD.55.3873",
    journal = "Phys. Rev. D",
    volume = "55",
    pages = "3873--3881",
    year = "1997"
}

@article{Cline:2011mm,
    author = "Cline, James M. and Kainulainen, Kimmo and Trott, Michael",
    title = "{Electroweak Baryogenesis in Two Higgs Doublet Models and B meson anomalies}",
    eprint = "1107.3559",
    archivePrefix = "arXiv",
    primaryClass = "hep-ph",
    doi = "10.1007/JHEP11(2011)089",
    journal = "JHEP",
    volume = "11",
    pages = "089",
    year = "2011"
}

@article{Branco:2011iw,
    author = "Branco, G. C. and Ferreira, P. M. and Lavoura, L. and Rebelo, M. N. and Sher, Marc and Silva, Joao P.",
    title = "{Theory and phenomenology of two-Higgs-doublet models}",
    eprint = "1106.0034",
    archivePrefix = "arXiv",
    primaryClass = "hep-ph",
    doi = "10.1016/j.physrep.2012.02.002",
    journal = "Phys. Rept.",
    volume = "516",
    pages = "1--102",
    year = "2012"
}

@article{Coleman:1973jx,
    author = "Coleman, Sidney R. and Weinberg, Erick J.",
    title = "{Radiative Corrections as the Origin of Spontaneous Symmetry Breaking}",
    doi = "10.1103/PhysRevD.7.1888",
    journal = "Phys. Rev. D",
    volume = "7",
    pages = "1888--1910",
    year = "1973"
}

@article{Ghorbani:2022vtv,
    author = "Ghorbani, Karim and Ghorbani, Parsa",
    title = "{W-boson mass anomaly from scale invariant 2HDM}",
    eprint = "2204.09001",
    archivePrefix = "arXiv",
    primaryClass = "hep-ph",
    doi = "10.1016/j.nuclphysb.2022.115980",
    journal = "Nucl. Phys. B",
    volume = "984",
    pages = "115980",
    year = "2022"
}

@article{Parwani:1991gq,
    author = "Parwani, Rajesh R.",
    title = "{Resummation in a hot scalar field theory}",
    eprint = "hep-ph/9204216",
    archivePrefix = "arXiv",
    reportNumber = "ITP-SB-91-64",
    doi = "10.1103/PhysRevD.45.4695",
    journal = "Phys. Rev. D",
    volume = "45",
    pages = "4695",
    year = "1992",
    note = "[Erratum: Phys.Rev.D 48, 5965 (1993)]"
}

@article{Becirevic:2015fmu,
    author = "Be\v{c}irevi\'c, D. and Bertuzzo, Enrico and Sumensari, Olcyr and Zukanovich Funchal, Renata",
    title = "{Can the new resonance at LHC be a CP-Odd Higgs boson?}",
    eprint = "1512.05623",
    archivePrefix = "arXiv",
    primaryClass = "hep-ph",
    reportNumber = "LPT-ORSAY-15-100",
    doi = "10.1016/j.physletb.2016.03.073",
    journal = "Phys. Lett. B",
    volume = "757",
    pages = "261--267",
    year = "2016"
}

@article{Gunion:2002zf,
    author = "Gunion, John F. and Haber, Howard E.",
    title = "{The CP conserving two Higgs doublet model: The Approach to the decoupling limit}",
    eprint = "hep-ph/0207010",
    archivePrefix = "arXiv",
    reportNumber = "SCIPP-02-10",
    doi = "10.1103/PhysRevD.67.075019",
    journal = "Phys. Rev. D",
    volume = "67",
    pages = "075019",
    year = "2003"
}

@article{ParticleDataGroup:2020ssz,
    author = "Zyla, P. A. and others",
    collaboration = "Particle Data Group",
    title = "{Review of Particle Physics}",
    doi = "10.1093/ptep/ptaa104",
    journal = "PTEP",
    volume = "2020",
    number = "8",
    pages = "083C01",
    year = "2020"
}

@article{Baak:2011ze,
    author = "Baak, M. and Goebel, M. and Haller, J. and Hoecker, A. and Kennedy, D. and Moenig, K. and Schott, M. and Stelzer, J.",
    collaboration = "Gfitter",
    title = "{Updated Status of the Global Electroweak Fit and Constraints on New Physics}",
    eprint = "1107.0975",
    archivePrefix = "arXiv",
    primaryClass = "hep-ph",
    reportNumber = "DESY-11-107, CERN-OPEN-2011-033, CERN-OPEN-2011-033, DESY-11-107",
    doi = "10.1140/epjc/s10052-012-2003-4",
    journal = "Eur. Phys. J. C",
    volume = "72",
    pages = "2003",
    year = "2012"
}

@article{Lu:2022bgw,
    author = "Lu, Chih-Ting and Wu, Lei and Wu, Yongcheng and Zhu, Bin",
    title = "{Electroweak precision fit and new physics in light of the W boson mass}",
    eprint = "2204.03796",
    archivePrefix = "arXiv",
    primaryClass = "hep-ph",
    doi = "10.1103/PhysRevD.106.035034",
    journal = "Phys. Rev. D",
    volume = "106",
    number = "3",
    pages = "035034",
    year = "2022"
}

@article{Peskin:1991sw,
    author = "Peskin, Michael E. and Takeuchi, Tatsu",
    title = "{Estimation of oblique electroweak corrections}",
    reportNumber = "SLAC-PUB-5618",
    doi = "10.1103/PhysRevD.46.381",
    journal = "Phys. Rev. D",
    volume = "46",
    pages = "381--409",
    year = "1992"
}

@article{Espinosa:2010hh,
    author = "Espinosa, Jose R. and Konstandin, Thomas and No, Jose M. and Servant, Geraldine",
    title = "{Energy Budget of Cosmological First-order Phase Transitions}",
    eprint = "1004.4187",
    archivePrefix = "arXiv",
    primaryClass = "hep-ph",
    reportNumber = "CERN-PH-TH-2010-027",
    doi = "10.1088/1475-7516/2010/06/028",
    journal = "JCAP",
    volume = "06",
    pages = "028",
    year = "2010"
}

@article{Grojean:2006bp,
    author = "Grojean, Christophe and Servant, Geraldine",
    title = "{Gravitational Waves from Phase Transitions at the Electroweak Scale and Beyond}",
    eprint = "hep-ph/0607107",
    archivePrefix = "arXiv",
    reportNumber = "CERN-PH-TH-2006-125",
    doi = "10.1103/PhysRevD.75.043507",
    journal = "Phys. Rev. D",
    volume = "75",
    pages = "043507",
    year = "2007"
}

@article{Arcadi:2022lpp,
    author = "Arcadi, Giorgio and Benincasa, Nico and Djouadi, Abdelhak and Kannike, Kristjan",
    title = "{Two-Higgs-doublet-plus-pseudoscalar model: Collider, dark matter, and gravitational wave signals}",
    eprint = "2212.14788",
    archivePrefix = "arXiv",
    primaryClass = "hep-ph",
    doi = "10.1103/PhysRevD.108.055010",
    journal = "Phys. Rev. D",
    volume = "108",
    number = "5",
    pages = "055010",
    year = "2023"
}

@article{Turner:1992tz,
    author = "Turner, Michael S. and Weinberg, Erick J. and Widrow, Lawrence M.",
    title = "{Bubble nucleation in first order inflation and other cosmological phase transitions}",
    reportNumber = "FERMILAB-PUB-91-334-A, CU-TP-558, IASSNS-HEP-92-21",
    doi = "10.1103/PhysRevD.46.2384",
    journal = "Phys. Rev. D",
    volume = "46",
    pages = "2384--2403",
    year = "1992"
}

@article{Ellis:2018mja,
    author = "Ellis, John and Lewicki, Marek and No, Jos\'e Miguel",
    title = "{On the Maximal Strength of a First-Order Electroweak Phase Transition and its Gravitational Wave Signal}",
    eprint = "1809.08242",
    archivePrefix = "arXiv",
    primaryClass = "hep-ph",
    reportNumber = "KCL-PH-TH/2018-46, CERN-TH/2018-197, IFT-UAM/CSIC-18-94, CERN-TH-2018-197",
    doi = "10.1088/1475-7516/2019/04/003",
    journal = "JCAP",
    volume = "04",
    pages = "003",
    year = "2019"
}

@article{Enqvist:1991xw,
    author = "Enqvist, K. and Ignatius, J. and Kajantie, K. and Rummukainen, K.",
    title = "{Nucleation and bubble growth in a first order cosmological electroweak phase transition}",
    reportNumber = "HU-TFT-91-35",
    doi = "10.1103/PhysRevD.45.3415",
    journal = "Phys. Rev. D",
    volume = "45",
    pages = "3415--3428",
    year = "1992"
}

@ARTICLE{1971AdPhy..20..325S,
       author = {{Shante}, Vinod K.~S. and {Kirkpatrick}, Scott},
        title = "{An introduction to percolation theory{\textdagger}}",
      journal = {Advances in Physics},
         year = 1971,
        month = may,
       volume = {20},
       number = {85},
        pages = {325-357},
          doi = {10.1080/00018737100101261},
       adsurl = {https://ui.adsabs.harvard.edu/abs/1971AdPhy..20..325S}
}

@ARTICLE{2017arXiv170200786A,
       author = {{Amaro-Seoane}, Pau and {Audley}, Heather and {Babak}, Stanislav and {Baker}, John and {Barausse}, Enrico and {Bender}, Peter and {Berti}, Emanuele and {Binetruy}, Pierre and {Born}, Michael and {Bortoluzzi}, Daniele and {Camp}, Jordan and {Caprini}, Chiara and {Cardoso}, Vitor and {Colpi}, Monica and {Conklin}, John and {Cornish}, Neil and {Cutler}, Curt and {Danzmann}, Karsten and {Dolesi}, Rita and {Ferraioli}, Luigi and {Ferroni}, Valerio and {Fitzsimons}, Ewan and {Gair}, Jonathan and {Gesa Bote}, Lluis and {Giardini}, Domenico and {Gibert}, Ferran and {Grimani}, Catia and {Halloin}, Hubert and {Heinzel}, Gerhard and {Hertog}, Thomas and {Hewitson}, Martin and {Holley-Bockelmann}, Kelly and {Hollington}, Daniel and {Hueller}, Mauro and {Inchauspe}, Henri and {Jetzer}, Philippe and {Karnesis}, Nikos and {Killow}, Christian and {Klein}, Antoine and {Klipstein}, Bill and {Korsakova}, Natalia and {Larson}, Shane L and {Livas}, Jeffrey and {Lloro}, Ivan and {Man}, Nary and {Mance}, Davor and {Martino}, Joseph and {Mateos}, Ignacio and {McKenzie}, Kirk and {McWilliams}, Sean T and {Miller}, Cole and {Mueller}, Guido and {Nardini}, Germano and {Nelemans}, Gijs and {Nofrarias}, Miquel and {Petiteau}, Antoine and {Pivato}, Paolo and {Plagnol}, Eric and {Porter}, Ed and {Reiche}, Jens and {Robertson}, David and {Robertson}, Norna and {Rossi}, Elena and {Russano}, Giuliana and {Schutz}, Bernard and {Sesana}, Alberto and {Shoemaker}, David and {Slutsky}, Jacob and {Sopuerta}, Carlos F. and {Sumner}, Tim and {Tamanini}, Nicola and {Thorpe}, Ira and {Troebs}, Michael and {Vallisneri}, Michele and {Vecchio}, Alberto and {Vetrugno}, Daniele and {Vitale}, Stefano and {Volonteri}, Marta and {Wanner}, Gudrun and {Ward}, Harry and {Wass}, Peter and {Weber}, William and {Ziemer}, John and {Zweifel}, Peter},
        title = "{Laser Interferometer Space Antenna}",
      journal = {arXiv e-prints},
         year = 2017,
        month = feb,
          eid = {arXiv:1702.00786},
        pages = {arXiv:1702.00786},
          doi = {10.48550/arXiv.1702.00786},
archivePrefix = {arXiv},
       eprint = {1702.00786},
 primaryClass = {astro-ph.IM},
       adsurl = {https://ui.adsabs.harvard.edu/abs/2017arXiv170200786A}
}

@article{Kawamura:2011zz,
    author = "Kawamura, Seiji and others",
    editor = "Buchman, Sasha and Sun, Ke-Xun",
    title = "{The Japanese space gravitational wave antenna: DECIGO}",
    doi = "10.1088/0264-9381/28/9/094011",
    journal = "Class. Quant. Grav.",
    volume = "28",
    pages = "094011",
    year = "2011"
}

@article{Corbin:2005ny,
    author = "Corbin, Vincent and Cornish, Neil J.",
    title = "{Detecting the cosmic gravitational wave background with the big bang observer}",
    eprint = "gr-qc/0512039",
    archivePrefix = "arXiv",
    doi = "10.1088/0264-9381/23/7/014",
    journal = "Class. Quant. Grav.",
    volume = "23",
    pages = "2435--2446",
    year = "2006"
}

@article{Hu:2017mde,
    author = "Hu, Wen-Rui and Wu, Yue-Liang",
    title = "{The Taiji Program in Space for gravitational wave physics and the nature of gravity}",
    doi = "10.1093/nsr/nwx116",
    journal = "Natl. Sci. Rev.",
    volume = "4",
    number = "5",
    pages = "685--686",
    year = "2017"
}

@article{Maggiore:1999vm,
    author = "Maggiore, Michele",
    title = "{Gravitational wave experiments and early universe cosmology}",
    eprint = "gr-qc/9909001",
    archivePrefix = "arXiv",
    reportNumber = "IFUP-TH-20-99",
    doi = "10.1016/S0370-1573(99)00102-7",
    journal = "Phys. Rept.",
    volume = "331",
    pages = "283--367",
    year = "2000"
}

@article{Kahniashvili:2021kls,
    author = "Kahniashvili, Tina and Clarke, Emma and Stepp, Jonathan and Brandenburg, Axel",
    title = "{Big Bang Nucleosynthesis Limits and Relic Gravitational Waves Detection Prospects}",
    eprint = "2111.09541",
    archivePrefix = "arXiv",
    primaryClass = "astro-ph.CO",
    reportNumber = "NORDITA-2021-089",
    doi = "10.1103/PhysRevLett.128.221301",
    journal = "Phys. Rev. Lett.",
    volume = "128",
    number = "22",
    pages = "221301",
    year = "2022"
}

@article{Bodeker:2009qy,
    author = "Bodeker, Dietrich and Moore, Guy D.",
    title = "{Can electroweak bubble walls run away?}",
    eprint = "0903.4099",
    archivePrefix = "arXiv",
    primaryClass = "hep-ph",
    doi = "10.1088/1475-7516/2009/05/009",
    journal = "JCAP",
    volume = "05",
    pages = "009",
    year = "2009"
}

@article{Guo:2020grp,
    author = "Guo, Huai-Ke and Sinha, Kuver and Vagie, Daniel and White, Graham",
    title = "{Phase Transitions in an Expanding Universe: Stochastic Gravitational Waves in Standard and Non-Standard Histories}",
    eprint = "2007.08537",
    archivePrefix = "arXiv",
    primaryClass = "hep-ph",
    doi = "10.1088/1475-7516/2021/01/001",
    journal = "JCAP",
    volume = "01",
    pages = "001",
    year = "2021"
}

@article{Ellis:2019oqb,
    author = "Ellis, John and Lewicki, Marek and No, Jos{\'e} Miguel and Vaskonen, Ville",
    title = "{Gravitational wave energy budget in strongly supercooled phase transitions}",
    eprint = "1903.09642",
    archivePrefix = "arXiv",
    primaryClass = "hep-ph",
    reportNumber = "KCL-PH-TH/2019-32, CERN-TH-2019-032, IFT-UAM/CSIC-19-32",
    doi = "10.1088/1475-7516/2019/06/024",
    journal = "JCAP",
    volume = "06",
    pages = "024",
    year = "2019"
}

@article{Caprini:2018mtu,
    author = "Caprini, Chiara and Figueroa, Daniel G.",
    title = "{Cosmological Backgrounds of Gravitational Waves}",
    eprint = "1801.04268",
    archivePrefix = "arXiv",
    primaryClass = "astro-ph.CO",
    doi = "10.1088/1361-6382/aac608",
    journal = "Class. Quant. Grav.",
    volume = "35",
    number = "16",
    pages = "163001",
    year = "2018"
}

@article{Gildener:1976ih,
    author = "Gildener, Eldad and Weinberg, Steven",
    title = "{Symmetry Breaking and Scalar Bosons}",
    reportNumber = "PRINT-76-0068 (HARVARD)",
    doi = "10.1103/PhysRevD.13.3333",
    journal = "Phys. Rev. D",
    volume = "13",
    pages = "3333",
    year = "1976"
}

@article{Dolan:1973qd,
    author = "Dolan, L. and Jackiw, R.",
    title = "{Symmetry Behavior at Finite Temperature}",
    reportNumber = "MIT-CTP-406",
    doi = "10.1103/PhysRevD.9.3320",
    journal = "Phys. Rev. D",
    volume = "9",
    pages = "3320--3341",
    year = "1974"
}

@article{Caprini:2024hue,
    author = "Caprini, Chiara and Jinno, Ryusuke and Lewicki, Marek and Madge, Eric and Merchand, Marco and Nardini, Germano and Pieroni, Mauro and Roper Pol, Alberto and Vaskonen, Ville",
    collaboration = "LISA Cosmology Working Group",
    title = "{Gravitational waves from first-order phase transitions in LISA: reconstruction pipeline and physics interpretation}",
    eprint = "2403.03723",
    archivePrefix = "arXiv",
    primaryClass = "astro-ph.CO",
    reportNumber = "LISA-COSWG-24-01, CERN-TH-2024-029",
    doi = "10.1088/1475-7516/2024/10/020",
    journal = "JCAP",
    volume = "10",
    pages = "020",
    year = "2024"
}

@article{Thrane:2013oya,
    author = "Thrane, Eric and Romano, Joseph D.",
    title = "{Sensitivity curves for searches for gravitational-wave backgrounds}",
    eprint = "1310.5300",
    archivePrefix = "arXiv",
    primaryClass = "astro-ph.IM",
    doi = "10.1103/PhysRevD.88.124032",
    journal = "Phys. Rev. D",
    volume = "88",
    number = "12",
    pages = "124032",
    year = "2013"
}

@misc{Bardeen:1995kv,
  author       = {Bardeen, William A.},
  title        = {On naturalness in the standard model},
  note         = {FERMILAB-CONF-95-391-T, presented at the Ontake Summer Institute on Particle Physics (1995)},
  year         = {1995}
}

@article{Meissner:2006zh,
    author = "Meissner, Krzysztof A. and Nicolai, Hermann",
    title = "{Conformal Symmetry and the Standard Model}",
    eprint = "hep-th/0612165",
    archivePrefix = "arXiv",
    doi = "10.1016/j.physletb.2007.03.023",
    journal = "Phys. Lett. B",
    volume = "648",
    pages = "312--317",
    year = "2007"
}

@article{Meissner:2007ru,
  author       = {Meissner, Krzysztof A. and Nicolai, Hermann},
  title        = {Effective action, conformal anomaly and the issue of quadratic divergences},
  journal      = {Phys. Lett. B},
  volume       = {660},
  pages        = {260--266},
  year         = {2008},
  doi          = {10.1016/j.physletb.2007.12.035},
  eprint       = {0710.2840},
  archivePrefix= {arXiv},
  primaryClass = {hep-th}
}

@article{Shaposhnikov:2008xi,
  author       = {Shaposhnikov, Mikhail and Zenhausern, Daniel},
  title        = {Quantum scale invariance, cosmological constant and hierarchy problem},
  journal      = {Phys. Lett. B},
  volume       = {671},
  number       = {1},
  pages        = {162--166},
  year         = {2009},
  doi          = {10.1016/j.physletb.2008.11.041},
  eprint       = {0809.3406},
  archivePrefix= {arXiv},
  primaryClass = {hep-th}
}

@article{Foot:2007as,
  author       = {Foot, Robert and Kobakhidze, Archil and McDonald, Kristian L. and Volkas, Raymond R.},
  title        = {A solution to the hierarchy problem from an almost decoupled hidden sector within a classically scale invariant theory},
  journal      = {Phys. Rev. D},
  volume       = {77},
  pages        = {035006},
  year         = {2008},
  doi          = {10.1103/PhysRevD.77.035006},
  eprint       = {0709.2750},
  archivePrefix= {arXiv},
  primaryClass = {hep-ph}
}

@article{Konstandin:2011dr,
  author       = {Konstandin, Thomas and Servant, Geraldine},
  title        = {Cosmological consequences of nearly conformal dynamics at the TeV scale},
  journal      = {JCAP},
  volume       = {12},
  pages        = {009},
  year         = {2011},
  doi          = {10.1088/1475-7516/2011/12/009},
  eprint       = {1104.4791},
  archivePrefix= {arXiv},
  primaryClass = {hep-ph}
}

@article{Jinno:2016knw,
  author       = {Jinno, Ryusuke and Takimoto, Masahiro},
  title        = {Probing classically conformal B-L model with gravitational waves},
  journal      = {Phys. Rev. D},
  volume       = {95},
  pages        = {015020},
  year         = {2017},
  doi          = {10.1103/PhysRevD.95.015020},
  eprint       = {1604.05035},
  archivePrefix= {arXiv},
  primaryClass = {hep-ph}
}

@article{Marzola:2017jzl,
  author       = {Marzola, Luca and Racioppi, Antonio and Vaskonen, Ville},
  title        = {Phase transition and gravitational wave phenomenology of scalar conformal extensions of the Standard Model},
  journal      = {Eur. Phys. J. C},
  volume       = {77},
  pages        = {484},
  year         = {2017},
  doi          = {10.1140/epjc/s10052-017-4996-1},
  eprint       = {1704.01034},
  archivePrefix= {arXiv},
  primaryClass = {hep-ph}
}

@article{Marzo:2018nov,
  author       = {Marzo, Carlo and Marzola, Luca and Vaskonen, Ville},
  title        = {Phase transition and vacuum stability in the classically conformal B-L model},
  journal      = {Eur. Phys. J. C},
  volume       = {79},
  pages        = {601},
  year         = {2019},
  doi          = {10.1140/epjc/s10052-019-7076-x},
  eprint       = {1811.11169},
  archivePrefix= {arXiv},
  primaryClass = {hep-ph}
}

@article{Iso:2017rjs,
  author       = {Iso, Satoshi and Serpico, Pasquale D. and Shimada, Kengo},
  title        = {QCD-Electroweak First-Order Phase Transition in a Supercooled Universe},
  journal      = {Phys. Rev. Lett.},
  volume       = {119},
  pages        = {141301},
  year         = {2017},
  doi          = {10.1103/PhysRevLett.119.141301},
  eprint       = {1704.04955},
  archivePrefix= {arXiv},
  primaryClass = {hep-ph}
}

@article{Lee:1973iz,
  author  = {Lee, T. D.},
  title   = {A Theory of Spontaneous T Violation},
  journal = {Phys. Rev. D},
  volume  = {8},
  pages   = {1226--1239},
  year    = {1973},
  doi     = {10.1103/PhysRevD.8.1226}
}

@article{Lee:1974fj,
  author  = {Lee, T. D.},
  title   = {C P nonconservation and spontaneous symmetry breaking},
  journal = {Phys. Rept.},
  volume  = {9},
  number  = {2},
  pages   = {143--177},
  year    = {1974},
  doi     = {10.1016/0370-1573(74)90020-9}
}

@article{Fayet:1976et,
  author  = {Fayet, P.},
  title   = {Supersymmetry and weak, electromagnetic and strong interactions},
  journal = {Phys. Lett. B},
  volume  = {64},
  number  = {2},
  pages   = {159--162},
  year    = {1976},
  doi     = {10.1016/0370-2693(76)90319-1}
}

@article{Fayet:1977yc,
  author  = {Fayet, P.},
  title   = {Spontaneously broken supersymmetric theories of weak, electromagnetic and strong interactions},
  journal = {Phys. Lett. B},
  volume  = {69},
  number  = {4},
  pages   = {489--494},
  year    = {1977},
  doi     = {10.1016/0370-2693(77)90852-8}
}

@article{Haber:1984rc,
  author  = {Haber, Howard E. and Kane, Gordon L.},
  title   = {The search for supersymmetry: Probing physics beyond the standard model},
  journal = {Phys. Rept.},
  volume  = {117},
  number  = {2-4},
  pages   = {75--263},
  year    = {1985},
  doi     = {10.1016/0370-1573(85)90051-1}
}

@misc{Guo:2025uhf,
  author        = {Guo, Xinyao and Miao, Haixing and Wang, Zhi-Wei and Yang, Huan and Zhou, Ye-Ling},
  title         = {There is Room at the Top: Fundamental Quantum Limits for Detecting Ultra-high Frequency Gravitational Waves},
  year          = {2025},
  eprint        = {2501.18146},
  archivePrefix = {arXiv},
  primaryClass  = {gr-qc},
  doi           = {10.48550/arXiv.2501.18146}
}

@article{Auclair:2022wcv,
  author        = {Auclair, Pierre and others},
  collaboration = {LISA Cosmology Working Group},
  title         = {Cosmology with the Laser Interferometer Space Antenna},
  journal       = {Living Rev. Relativ.},
  volume        = {26},
  number        = {1},
  pages         = {5},
  year          = {2023},
  doi           = {10.1007/s41114-023-00045-2},
  eprint        = {2204.05434},
  archivePrefix = {arXiv},
  primaryClass  = {astro-ph.CO}
}

@article{Arzoumanian:2021fsy,
  author        = {Arzoumanian, Zaven and others},
  collaboration = {NANOGrav Collaboration},
  title         = {Searching for Gravitational Waves from Cosmological Phase Transitions with the NANOGrav 12.5-Year Dataset},
  journal       = {Phys. Rev. Lett.},
  volume        = {127},
  number        = {25},
  pages         = {251302},
  year          = {2021},
  doi           = {10.1103/PhysRevLett.127.251302},
  eprint        = {2104.13930},
  archivePrefix = {arXiv},
  primaryClass  = {astro-ph.CO}
}

@article{Xue:2021gyq,
  author  = {Xue, Xiao and Bian, Ligong and Shu, Jing and Yuan, Qiang and Zhu, Xingjiang and others},
  title   = {Constraining Cosmological Phase Transitions with the Parkes Pulsar Timing Array},
  journal = {Phys. Rev. Lett.},
  volume  = {127},
  number  = {25},
  pages   = {251303},
  year    = {2021},
  doi     = {10.1103/PhysRevLett.127.251303}
}

@article{Aggarwal:2021icw,
  author  = {Aggarwal, Nancy and others},
  title   = {Challenges and opportunities of gravitational-wave searches at MHz to GHz frequencies},
  journal = {Living Rev. Relativ.},
  volume  = {24},
  number  = {1},
  pages   = {4},
  year    = {2021},
  doi     = {10.1007/s41114-021-00032-5}
}

@article{Huang:2020bbe,
    author = "Huang, Wei-Chih and Sannino, Francesco and Wang, Zhi-Wei",
    title = "{Gravitational Waves from Pati-Salam Dynamics}",
    eprint = "2004.02332",
    archivePrefix = "arXiv",
    primaryClass = "hep-ph",
    reportNumber = "CP3-Origins-2020-05 DNRF90",
    doi = "10.1103/PhysRevD.102.095025",
    journal = "Phys. Rev. D",
    volume = "102",
    number = "9",
    pages = "095025",
    year = "2020"
}

@article{Steele:2012av,
    author = "Steele, T. G. and Wang, Zhi-Wei",
    title = "{Is Radiative Electroweak Symmetry Breaking Consistent with a 125 GeV Higgs Mass?}",
    eprint = "1209.5416",
    archivePrefix = "arXiv",
    primaryClass = "hep-ph",
    doi = "10.1103/PhysRevLett.110.151601",
    journal = "Phys. Rev. Lett.",
    volume = "110",
    number = "15",
    pages = "151601",
    year = "2013"
}

@article{Steele:2013fka,
    author = "Steele, T. G. and Wang, Zhi-Wei and Contreras, D. and Mann, R. B.",
    title = "{Viable dark matter via radiative symmetry breaking in a scalar singlet Higgs portal extension of the standard model}",
    eprint = "1310.1960",
    archivePrefix = "arXiv",
    primaryClass = "hep-ph",
    doi = "10.1103/PhysRevLett.112.171602",
    journal = "Phys. Rev. Lett.",
    volume = "112",
    number = "17",
    pages = "171602",
    year = "2014"
}

@article{Steele:2014dsa,
    author = "Steele, T. G. and Wang, Zhi-Wei and McKeon, D. G. C.",
    title = "{Multiscale renormalization group methods for effective potentials with multiple scalar fields}",
    eprint = "1409.3489",
    archivePrefix = "arXiv",
    primaryClass = "hep-ph",
    doi = "10.1103/PhysRevD.90.105012",
    journal = "Phys. Rev. D",
    volume = "90",
    number = "10",
    pages = "105012",
    year = "2014"
}

@article{Hansen:2017pwe,
    author = "Hansen, Frederik F. and Janowski, Tadeusz and Lang{\ae}ble, Kasper and Mann, Robert B. and Sannino, Francesco and Steele, Tom G. and Wang, Zhi-Wei",
    title = "{Phase structure of complete asymptotically free SU($N_c$) theories with quarks and scalar quarks}",
    eprint = "1706.06402",
    archivePrefix = "arXiv",
    primaryClass = "hep-ph",
    reportNumber = "CP3-ORIGINS-2017-022, CERN-TH-2017-133",
    doi = "10.1103/PhysRevD.97.065014",
    journal = "Phys. Rev. D",
    volume = "97",
    number = "6",
    pages = "065014",
    year = "2018"
}

@article{Hill:2014mqa,
    author = "Hill, Christopher T.",
    title = "{Is the Higgs Boson Associated with Coleman-Weinberg Dynamical Symmetry Breaking?}",
    eprint = "1401.4185",
    archivePrefix = "arXiv",
    primaryClass = "hep-ph",
    reportNumber = "FERMILAB-PUB-14-003-T",
    doi = "10.1103/PhysRevD.89.073003",
    journal = "Phys. Rev. D",
    volume = "89",
    number = "7",
    pages = "073003",
    year = "2014"
}

@article{Chataignier:2018kay,
    author = "Chataignier, Leonardo and Prokopec, Tomislav and Schmidt, Michael G. and {\'S}wie{\.z}ewska, Bogumi{\l}a",
    title = "{Systematic analysis of radiative symmetry breaking in models with extended scalar sector}",
    eprint = "1805.09292",
    archivePrefix = "arXiv",
    primaryClass = "hep-ph",
    doi = "10.1007/JHEP08(2018)083",
    journal = "JHEP",
    volume = "08",
    pages = "083",
    year = "2018"
}

@article{Bittar:2025lcr,
    author = "Bittar, Pedro and Roy, Subhojit and Wagner, Carlos E. M.",
    title = "{Self consistent thermal resummation: a case study of the phase transition in 2HDM}",
    eprint = "2504.02024",
    archivePrefix = "arXiv",
    primaryClass = "hep-ph",
    reportNumber = "EFI-25-3",
    doi = "10.1007/JHEP12(2025)021",
    journal = "JHEP",
    volume = "12",
    pages = "021",
    year = "2025"
}

@article{Ferreira:2004yd,
    author = "Ferreira, P. M. and Santos, R. and Barroso, A.",
    title = "{Stability of the tree-level vacuum in two Higgs doublet models against charge or CP spontaneous violation}",
    eprint = "hep-ph/0406231",
    archivePrefix = "arXiv",
    doi = "10.1016/j.physletb.2004.10.022",
    journal = "Phys. Lett. B",
    volume = "603",
    pages = "219--229",
    year = "2004",
    note = "[Erratum: Phys. Lett. B 629, 114 (2005)]"
}

@article{Witten:1980ez,
    author = "Witten, Edward",
    title = "{Cosmological Consequences of a Light Higgs Boson}",
    doi = "10.1016/0550-3213(81)90182-6",
    journal = "Nucl. Phys. B",
    volume = "177",
    pages = "477--488",
    year = "1981"
}

@article{Dine:1992wr,
    author = "Dine, Michael and Leigh, Robert G. and Huet, Patrick Y. and Linde, Andrei D. and Linde, Dmitri A.",
    title = "{Towards the theory of the electroweak phase transition}",
    eprint = "hep-ph/9203203",
    archivePrefix = "arXiv",
    doi = "10.1103/PhysRevD.46.550",
    journal = "Phys. Rev. D",
    volume = "46",
    pages = "550--571",
    year = "1992"
}

@article{Alanne:2019bsm,
    author = "Alanne, Tommi and Hugle, Thomas and Platscher, Moritz and Schmitz, Kai",
    title = "{A fresh look at the gravitational-wave signal from cosmological phase transitions}",
    eprint = "1909.11356",
    archivePrefix = "arXiv",
    primaryClass = "hep-ph",
    doi = "10.1007/JHEP03(2020)004",
    journal = "JHEP",
    volume = "03",
    pages = "004",
    year = "2020"
}

@article{Athron:2022mmm,
    author = "Athron, Peter and Bal\'azs, Csaba and Morris, Lachlan",
    title = "{Supercool subtleties of cosmological phase transitions}",
    eprint = "2212.07559",
    archivePrefix = "arXiv",
    primaryClass = "hep-ph",
    doi = "10.1088/1475-7516/2023/03/006",
    journal = "JCAP",
    volume = "03",
    pages = "006",
    year = "2023"
}

@article{Salvio:2023qgb,
    author = "Salvio, Alberto",
    title = "{Model-Independent Radiative Symmetry Breaking and Gravitational Waves}",
    eprint = "2302.10212",
    archivePrefix = "arXiv",
    primaryClass = "hep-ph",
    doi = "10.1088/1475-7516/2023/04/051",
    journal = "JCAP",
    volume = "04",
    pages = "051",
    year = "2023"
}

@article{Salvio:2023ynn,
    author = "Salvio, Alberto",
    title = "{Supercooling in Radiative Symmetry Breaking: Theory Extensions, Gravitational Wave Detection and Primordial Black Holes}",
    eprint = "2307.04694",
    archivePrefix = "arXiv",
    primaryClass = "hep-ph",
    doi = "10.1088/1475-7516/2023/12/046",
    journal = "JCAP",
    volume = "12",
    pages = "046",
    year = "2023"
}

@article{Levi:2022bzt,
    author = "Levi, Noam and Opferkuch, Toby and Redigolo, Diego",
    title = "{The supercooling window at weak and strong coupling}",
    eprint = "2212.08085",
    archivePrefix = "arXiv",
    primaryClass = "hep-ph",
    doi = "10.1007/JHEP02(2023)125",
    journal = "JHEP",
    volume = "02",
    pages = "125",
    year = "2023"
}

\end{document}